\documentclass[]{article}

\usepackage[margin=1.3in]{geometry}

\usepackage{amsmath,amsthm,amssymb,euscript,mathtools} % AMS-LaTeX  
\usepackage{bm} % bold math

\usepackage{graphicx,color,ae,fancyvrb,caption}
\usepackage[T1]{fontenc}

\usepackage{rotating,epstopdf}

\usepackage{setspace}
\usepackage{lineno}

\usepackage{array}
\usepackage{booktabs}
\usepackage{dcolumn}
\usepackage{threeparttable}
\usepackage{siunitx}
\usepackage{natbib}
\usepackage[colorlinks=true,allcolors=blue]{hyperref}

\let\cite\citep

\author{%
  Tobias R\"{u}ttenauer $^\dag$ \footnotemark[1]%
  \and Ozan Aksoy $^\dag$%
  }

\date{$^\dag$ UCL Social Research Institute, University College London, 55-59 Gordon Square, London WC1H 0NU, UK}

\title{When Can We Use Two-Way Fixed-Effects (TWFE): A Comparison of TWFE and Novel Dynamic Difference-in-Differences Estimators}

\begin{document}

\maketitle
{\renewcommand{\thefootnote}{\fnsymbol{footnote}}
\footnotetext[1]{Email: t.ruttenauer@ucl.ac.uk}
}

\begin{abstract}

The conventional Two-Way Fixed Effects (TWFE) estimator has recently come under scrutiny, particularly in the context of staggered treatment adoption. A growing body of literature has shown that TWFE can yield biased estimates when treatment effects are heterogeneous across time or groups. In response, several advanced dynamic Difference-in-Differences (DiD) estimators have been developed to address this issue. However, confusion remains in applied research regarding when TWFE is biased, what problems the new estimators solve, and what limitations they still face. 
In this study, we provide an accessible overview of the recent DiD literature and clarify the conditions under which TWFE becomes inconsistent. Using Monte Carlo simulations, we evaluate the performance of TWFE and five dynamic DiD estimators across a range of realistic scenarios, including violations of key identifying assumptions such as parallel trends and no anticipation. Our results show that large parts of the bias in TWFE can be avoided by using an event-time specification. Under time- and group-heterogeneity, however, TWFE is clearly outperformed by the new dynamic DiD estimators, which are designed to capture these heterogeneous treatment effects. All estimators, however, are sensitive to anticipation effects and particularly to violations of the parallel trends assumption. Some of the new estimators are more sensitive to the former while others are more sensitive to the latter.
We conclude that no estimator is universally superior. Instead, each involves trade-offs depending on the data structure and the plausibility of identifying assumptions. Based on our findings, we offer practical guidance for applied researchers on when and how to use TWFE and the newer dynamic DiD estimators.

\bigskip\noindent\textbf{Keywords:} Difference-in-Differences; Monte Carlo Simulation; Two-Way Fixed-Effects; Dynamic Treatment Effects; Treatment Effect Heterogeneity; Parallel-Trends

\end{abstract}

\vfill

\begin{footnotesize}
Acknowledgements: An earlier version of this study was presented at the UCL Centre for Quantitative Social Science, RC28 in Milan, and the Life Trajectories Workshop in Cologne. We particularly thank Josef Brüderl, Alex Bryson, Lorraine Dearden, Pablo Geraldo, Hedvig Horvath, Fabian Kratz, and Burak Sonmez for their valuable comments. 
%A replication package with the simulation and analysis code is available on the author's \href{https://github.com/ruettenauer/did_sim}{Github repository}.
\end{footnotesize}

\thispagestyle{empty}
\clearpage
\onehalfspacing

\section{Introduction}

Researchers in the social sciences are often interested in causal questions. To estimate causal effects, they frequently rely on comparing the change in outcomes before and after a treatment to the change observed over the same period in a control group that did not receive the treatment. This approach is known as difference-in-differences (DiD) analysis. DiD estimators are among the most widely used tools in the social sciences \cite{Angrist.2015, Huntington-Klein.2021}. Because DiD compares within-unit changes over time between treated and untreated units, it offers strong leverage to control for both measured and unmeasured unit-specific factors that are constant over time \cite{Wooldridge.2010}. When extended to multiple time periods, DiD designs are often implemented using a Two-Way Fixed Effects (TWFE) regression specification. In the canonical 2$\times$2 DiD design, where individual observations $i$ belong to either a treatment group ($s = 1$) or a control group ($s = 0$), and there are two time periods $t = \{0, 1\}$ (pre- and post-treatment), the TWFE model is specified as:

\begin{equation}
\label{eqn:twfe_did}
y_{ist} = \beta_{TWFE} D_{st} + \alpha_s + \zeta_t + \epsilon_{it}
\end{equation}

\noindent where the outcome $y_{ist}$ is regressed on group fixed effects $\alpha_s$ (indicating treatment group membership), time fixed effects $\zeta_t$ (indicating whether the period is after the treatment), and a binary treatment indicator $D_{it}$ that switches from 0 to 1 when unit $i$ receives the treatment. For simplicity, we omit time-varying covariates. In this basic 2$\times$2 setup, the coefficient $\beta_{TWFE}$ in Equation~\ref{eqn:twfe_did} captures an intuitive estimand: the average treatment effect on the treated (ATT) \cite{Wooldridge.2010}. That is,

\begin{equation}
\label{eqn:did}
\beta_{TWFE} = \mathrm{E}(\Delta y_{T}) - \mathrm{E}(\Delta y_{C}) = [\mathrm{E}(y_{T}^{post}) - \mathrm{E}(y_{T}^{pre})] - [\mathrm{E}(y_{C}^{post}) - \mathrm{E}(y_{C}^{pre})].
\end{equation}

In such a 2$\times$2 setup, DiD and $\beta_{TWFE}$ provide a consistent estimate of the average causal treatment effect on the treated under the assumptions of parallel trends and no anticipation. Due to its intuitive interpretation and its strengths in estimating causal effects, TWFE has become popular in sociology and the broader social sciences. As \citet[][p.~43]{Gangl.2010} notes: ``it is hard to overstate the gain in identifying power provided by the beautifully simple method of FE estimation over standard cross-sectional estimators [and] the appeal of FE methods has only been growing over the past decade as panel data have increasingly become available." Indeed, TWFE has been widely applied to a range of topics, including the effects of demographic events on labour market outcomes \cite{Bruderl.2019, Schechtl.2023, Struffolino.2023, Zoch.2023}, the impact of policy measures or political parties \cite{Currie.2023, Dochow.2021, Kneip.2009, Aksoy.2021}, and the influence of environmental shocks \cite{Currie.2015, Osberghaus.2022, Ruttenauer.2024b}.

While TWFE performs well in the simple two-period, two-group setup, its interpretation becomes less clear in \emph{staggered} designs -- where treatment is rolled out over time across units, such that some units are treated early, others later, and some never. In such cases, it is not immediately obvious what the TWFE estimator captures. As \citet[][p.~33]{Wooldridge.2021} notes, ``for staggered interventions, the basic TWFE estimator has come under considerable scrutiny lately.'' Recent econometric research has highlighted several issues with TWFE in staggered designs \cite{Goodman-Bacon.2021, Callaway.2020, Sun.2021, Wooldridge.2021, DeChaisemartin.2020}. These concerns stem from the realisation that $\beta_{TWFE}$ in Equation~\ref{eqn:twfe_did} is, in fact, a non-intuitive weighted average of multiple contrasts between groups experiencing different treatment trajectories \cite{Goodman-Bacon.2021}. This can lead to biased estimates when treatment effects are heterogeneous across groups or over time.

To address these issues, researchers have proposed several alternative estimators that are robust to treatment effect heterogeneity. These include methods that estimate and aggregate multiple DiD contrasts \cite{Callaway.2020, Sun.2021}, imputation-based estimators that construct counterfactuals for treated units \cite{Borusyak.2024}, matrix completion approaches that combine DiD with synthetic control methods \cite{Athey.2021}, and extended TWFE models with more flexible parametrisations \cite{Wooldridge.2021}. Several recent reviews provide technical overviews of these estimators and their assumptions \cite[e.g.,][]{Freedman.2023, Roth.2023}.

However, we argue that this debate has led to some misjudgement of conventional TWFE estimators in applied research. As \citet{Wooldridge.2021} emphasises, ``there is nothing inherently wrong with using TWFE'' (p.~68), provided that researchers allow for sufficient treatment heterogeneity in their model specification, for example, by including a flexible event-time function. Moreover, there appears to be some confusion -- particularly outside the econometrics literature -- regarding the capabilities and limitations of the new DiD estimators. While these methods address specific issues related to treatment effect heterogeneity, their consistency still relies on assumptions such as parallel trends. A recent survey in political science \cite{Chiu.2023} found that most published results remain qualitatively unchanged when comparing TWFE with imputation-based methods. Violations of the parallel trends assumption appear to be a more consequential than issues of treatment effect heterogeneity.

This study has three main objectives. First, we aim to introduce the recent staggered DiD literature to social scientists in an accessible manner. To this end, we begin by presenting the conventional TWFE estimator and the problems identified in the recent econometric literature. We then explain the newly proposed alternative estimators in an intuitive way. Second, we demonstrate the behaviour of TWFE and alternative dynamic DiD estimators in an applied panel data setting. We use Monte Carlo simulations based on a large $N$ and a staggered design in which treatment is distributed across the whole observation period -- unlike macroeconomic settings where treatment occurs in only a few time periods. In our simulations, we gradually introduce realistic deviations from the ideal conditions for which the estimators were designed. We then assess how robust each estimator (including TWFE) is to these violations. This allows us to provide practical insights into the performance of different estimators in applied research.\footnote{The code is available on the author's GitHub repository [Link will be added].} Finally, we offer recommendations for best practices in data analysis.

\section{Two-Way Fixed Effects for Staggered Treatments}

In a multi-period setting with $T$ time periods and multiple treatment timings (e.g., marriage or labour market entry), researchers often rely on TWFE by including individual-specific fixed effects $\alpha_i$ and time-specific fixed effects $\zeta_t$, rather than group-specific indicators as in Equation~\ref{eqn:twfe_did}:

\begin{equation}
\label{eqn:twfe}
y_{it} = \beta_{TWFE} D_{it} + \alpha_i + \zeta_t + \epsilon_{it}.
\end{equation}

The coefficient $\beta_{TWFE}$ identifies the average treatment effect on the treated (ATT) under the strict exogeneity assumption: $\mathrm{E}(\epsilon_{it}|x_{i1}, \ldots, x_{iT}, \alpha_i) = 0 ~ \mathrm{for}~ t = 1,2,\ldots,T$ \cite{Wooldridge.2010}. This implies that the idiosyncratic errors $\epsilon_{it}$ are uncorrelated with the covariates in all periods (past and future) conditional on the unobserved effect $\alpha_i$. This assumption entails: (a) parallel trends (in the absence of treatment, the difference between treated and untreated units remains constant over time), (b) no anticipation (units do not respond to treatment before receiving it), (c) no unobserved time-varying confounders, (d) no carry-over effects (past treatment does not affect current outcomes), and (e) no feedback (past outcomes do not affect current treatment). 

As noted earlier, in the canonical two-period, two-group case, the TWFE model in Equation~\ref{eqn:twfe} yields a highly intuitive quantity: the $2 \times 2$ DiD estimator, identical to Equation~\ref{eqn:twfe_did}. Despite its simplicity and interpretability, this $2 \times 2$ DiD relies on weaker assumptions, requiring only (a) parallel trends and (b) no anticipation for consistent estimation of the ATT \cite{Ghanem.2024, Roth.2023}.

When treatment is assigned at varying time points -- so that some units are treated early, others later, and some never -- we refer to this as a \emph{staggered} design. In such settings, \citet{Goodman-Bacon.2021} shows that the $\beta_{TWFE}$ coefficient in Equation~\ref{eqn:twfe} is effectively a weighted average of multiple $2 \times 2$ DiD estimators, each comparing groups with different treatment timings. \citet{Goodman-Bacon.2021} derives the exact weights for each of these $2 \times 2$ DiD comparisons that collectively form $\beta_{TWFE}$. These weights are proportional to the number of observations in each comparison and the variance of the treatment indicator within each pair. A comparison pair consists of a treatment group that receives treatment at a specific time and a control group that does not receive treatment during the relevant window.

\begin{figure}[t]
  \centering
  \includegraphics[width=0.8\linewidth]{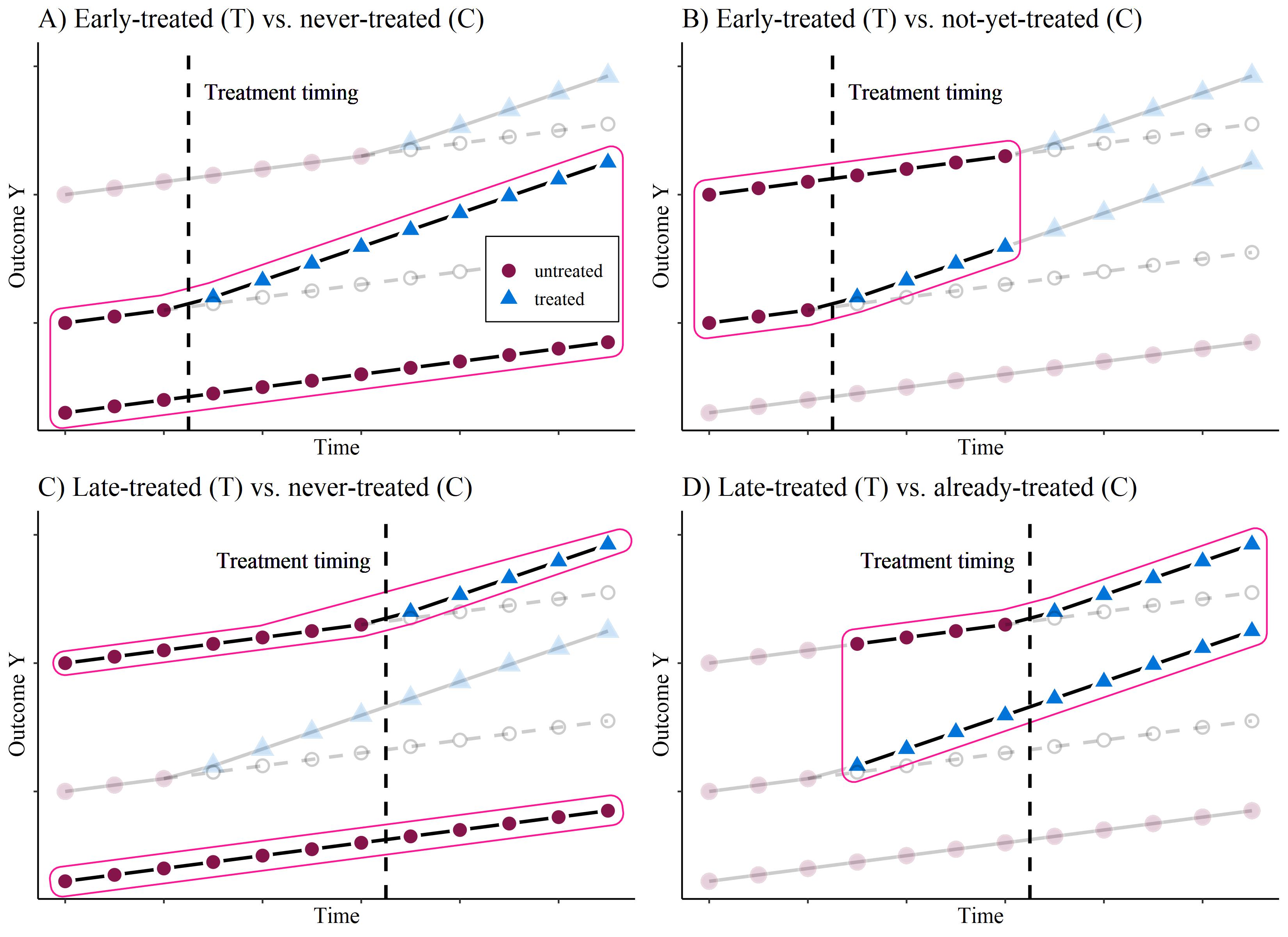}
  \caption{The four contrasting pairs of all 2-group DiD estimators that go into $\beta_{TWFE}$ in a TWFE regression with staggered treatment roll out, in this case with three groups: never-treated group, early-treated, and late-treated groups. Blue triangles are observations under treatment while red dots are untreated observations. Dashed horizontal lines show the counterfactual trajectory. Panel A) compares the early-treated to the never-treated, panel B) compares early-treated to the not-yet-treated observations of the late-treated, panel C) compares the late-treated to the never-treated, and panel D) compares the late-treated (as treatment group) to the already-treated of the early-treated (as control group).
  }
  \label{fig:Bacon}
\end{figure}

Figure~\ref{fig:Bacon} illustrates examples of the $2 \times 2$ DiD comparisons that contribute to $\beta_{TWFE}$. In panels A and C, treated units are compared to never-treated units: panel A shows early-treated vs. never-treated, and panel C shows late-treated vs. never-treated. In panel B, the early-treated group is compared to the late-treated group before the latter receives treatment (not-yet-treated). Some of those contrasts receive higher weights in $\beta_{TWFE}$ because of the timing of treatment but these comparisons are generally unproblematic.

However, panel D presents a problematic comparison: the late-treated group is compared to the early-treated group \emph{after} the latter has already received treatment (already-treated). This is the so-called ``forbidden comparison'' \cite{Borusyak.2024}, which introduces bias into $\beta_{TWFE}$.

In the stylised example of panel D, the issue arises because the ``control group'' (early-treated) is already experiencing an ongoing treatment effect, which biases the estimated treatment effect for this contrast. Although all groups follow parallel trends before treatment, the treated units deviate from the counterfactual trend (represented by empty circles) after treatment. This results in a downward bias in $\beta_{TWFE}$. Notably, this bias occurs even when treatment effects are identical across early and late-treated groups. If the early-treated group experienced a stronger effect, the bias could be even more severe -- potentially reversing the sign of the estimated effect if the ``control group'' overtakes the treated group. Depending on the direction and timing of treatment effects, as well as group-specific heterogeneity, the bias can vary in magnitude and direction, and in extreme cases, may even flip signs \cite{Roth.2023}.

By contrast, if the treatment effect follows a simple step-function (a one-time increase), the outcome trajectory of the already-treated group remains parallel to the counterfactual trajectory after treatment (see Figure~\ref{fig:step}). If there is no systematic variation of the treatment effect according to treatment timing, the comparison does not induce bias in case of a step-function treatment effect. The key issue with the ``forbidden comparison'' is that ongoing treatment effects cause the already-treated group’s trajectory to diverge from the counterfactual \cite{Borusyak.2024, Callaway.2020, DeChaisemartin.2020, Sun.2021}.

To summarise, the recent criticism of TWFE applies specifically to settings with time-varying (dynamic) treatment effects. In such cases, a conventional TWFE estimator with a single treatment indicator ($0 =$ not treated, $1 =$ treated) and a single coefficient will be biased due to the inclusion of ``forbidden comparisons''. However, if treatment effects are constant over time, the TWFE estimator still identifies the (variance-weighted) average treatment effect on the treated, even in staggered designs. When treatment effects vary across units but remain constant over time -- such as a step-level increase -- units treated mid-period may be overrepresented in $\beta_{TWFE}$, but the estimator remains consistent. Thus, the shortcomings of TWFE are not inherent to the estimator itself, but rather stem from misspecification of the treatment effect's functional form \cite{Goodman-Bacon.2021, Wooldridge.2021}. As we will show below, when the dynamic nature of treatment effects is correctly modelled using a flexible event-time specification -- i.e., a set of dichotomous indicators starting from the treatment onset \cite{Ludwig.2021} -- TWFE can again recover the time-specific ATT.

\section{Bias in TWFE}

In the previous section, we outlined the intuition behind the bias in TWFE estimators when treatment effects vary across time and groups. We did so by adapting the original example from \cite{Goodman-Bacon.2021}. In this stylised scenario, there are two treatment-timing groups, a clear separation between early and late treatment groups with potentially different treatment effects, and a trend-breaking treatment effect that persists indefinitely \cite{Callaway.2020, Goodman-Bacon.2021}.

However, in many sociological applications, we rely on individual-level panel data (large $N$, small $T$), and treatment effects tend to be short-lived. These effects often build up gradually after treatment and then fade out, resulting in an inverted-U shape. Such short-lived effects are empirically observed across various domains, from the impact of life events on happiness \cite{Bernardi.2017, Kratz.2020a, Clark.2013} to the social consequences of environmental shocks \cite{Baccini.2021, ConteKeivabu.2022, Currie.2015}.

\begin{figure}[t]
    \centering
    \includegraphics[width=\textwidth]{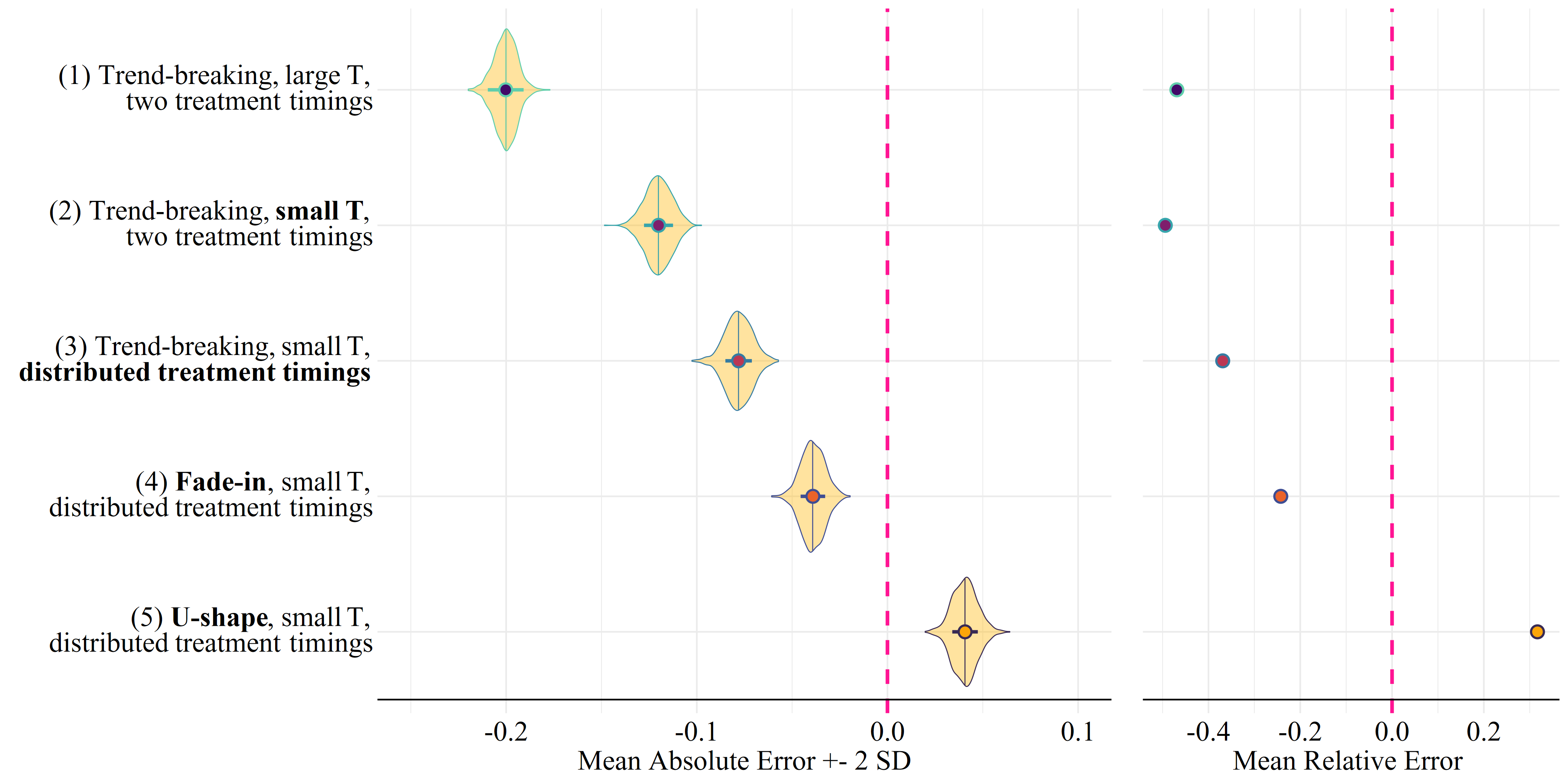}
    \caption{Monte Carlo results for a single summary measure of the effect estimates. Shown are the average absolute biases +/- two standard deviations on the left, and the relative biases on the right. Across all setups, the number of observations is held constant at $N = 1{,}000$. Large T refers to 30 time periods while small T 15. The yellow violin plots depict the distribution of the 1{,}000 individual estimates. The pink dashed line indicates the true effect (zero bias).}
    \label{fig:setup} 
\end{figure}

In Figure \ref{fig:setup}, we present empirical results on the magnitude of TWFE bias across several settings. For methodological details, please refer to the methods section. At the top of Figure \ref{fig:setup}, we begin with the stylised example described above: trend-breaking treatment effects, two treatment timings, a sharp distinction between groups with different treatment effects, and a long panel (large $T$). We then gradually transition to set-ups that resemble individual-level panel data. The absolute bias is shown on the left, and the relative bias on the right.

In the first stylised scenario (1), we observe an average bias of -0.2, corresponding to a relative bias of approximately 47\% -- a substantial distortion. When the number of time periods is reduced from 30 to 15 (small $T$), the absolute error nearly halves, while the relative error remains unchanged. This occurs because the magnitude of the true treatment effect diminishes in a trend-breaking setting as the number of time periods decreases. However, two factors influence both absolute and relative bias: (a) When we move from a two-timing treatment group setup to one where treatment timings are (normally) distributed across the observation window (e.g., marriage in an individual-level panel), the bias decreases (scenario 3); (b) When we shift from a trend-breaking to a fading (or inverse U-shaped) treatment effect, the bias also decreases (scenario 4). Switching to an inverse-U shaped treatment effect in scenario (5) changes the sign of the bias and slightly increases the relative bias to 32\%. 
%While lower than in the trend-breaking case, the magnitude of the bias in TWFE under heterogeneous treatment groups and dynamic treatment patters remains substantial.

Compared to the default stylised setting commonly used to demonstrate the issues with the TWFE, the bias due to heterogeneous treatment effects is somewhat lower in scenarios with smaller $T$, fading or U-shaped treatment effects, and treatment timings distributed over the observation period. Nevertheless, the remaining bias can still be substantial. It is therefore advisable -- in both macroeconomic and sociological panel settings -- to account for the possibility of time-varying treatment effects. In the following section, we compare the performance of several novel dynamic DiD estimators and conventional event-time TWFE specifications in addressing this issue.

\section{Dynamic DiD estimators}

There is now a growing and diverse set of alternative estimators designed to address the issues associated with TWFE in staggered treatment settings \cite[for formal reviews, see][]{Chiu.2023, Roth.2023}. In this study, we focus on five such alternatives, each representing a distinct methodological approach to mitigating the limitations of TWFE.

The first approach decomposes the overall effect into multiple $2 \times 2$ DiD comparisons, explicitly excluding the ``forbidden comparisons'' between early- and late-treated groups. These individual estimates are then aggregated into a weighted average that meaningfully summarises the treatment effect. The estimators proposed by \citet{Sun.2021} and \citet{Callaway.2020} fall under this category. The second approach imputes the missing counterfactual outcomes for treated units after treatment using a regression-based model \cite{Borusyak.2024}. The third approach also imputes counterfactuals but does so using a different technique: matrix completion, a method originally developed in computer science for recovering missing data \cite{Athey.2021}. The fourth approach involves estimating an extended version of TWFE (ETWFE), which allows treatment effects to vary across groups and over time by including appropriate interaction terms \cite{Wooldridge.2021}. In the following sections, we describe each of these approaches in more detail.

\subsection{Disaggregation-Based Estimators}

This class of estimators begins by decomposing the target parameter into multiple $2 \times 2$ DiD contrasts. The number of such contrasts depends on the specific timing of treatment across groups. These multiple comparisons can then be aggregated into a single summary estimate.

\citet{Callaway.2020} define the ATT at time point $t$ for a group that first received treatment in period $g$ as $\delta_{g,t}$. This ATT depends on both $g$ and $t$, meaning that each group treated at a different time has its own time-specific effect -- so the treatment effect for each group $g$ may vary over time. For example, $\delta_{5,10}$ denotes the average treatment effect at time $t = 10$ for the group first treated at $t = 5$. They further define a group- and time-specific difference-in-differences as:

\begin{equation}
\label{eqn:callaway1}
\delta_{g,t} = E(\Delta y_{g}) - E(\Delta y_{C}) = [E(y_{g}^{t}) - E(y_{g}^{g-1})] - [E(y_{C}^{t}) - E(y_{C}^{g-1})].
\end{equation}

Here, the term $[E(y_{g}^{t}) - E(y_{g}^{g-1})]$ captures the change in the average outcome for group $g$ from the pre-treatment period $g - 1$ to time $t$. In principle, any pre-treatment period $g - n$ (with $1 \leq n < g$) could be used as the reference. The term $[E(y_{C}^{t}) - E(y_{C}^{g-1})]$ captures the same change for a control group. \citet{Callaway.2020} propose two options for defining the control group: (1) the never-treated units ($[E(y_{\infty}^{t}) - E(y_{\infty}^{g-1})]$), and (2) the not-yet-treated units ($[E(y_{g'}^{t}) - E(y_{g'}^{g-1})|g'>t]$). These can also be combined into a joint control group.

\begin{figure}[t]
  \centering
  \includegraphics[width=0.9\linewidth]{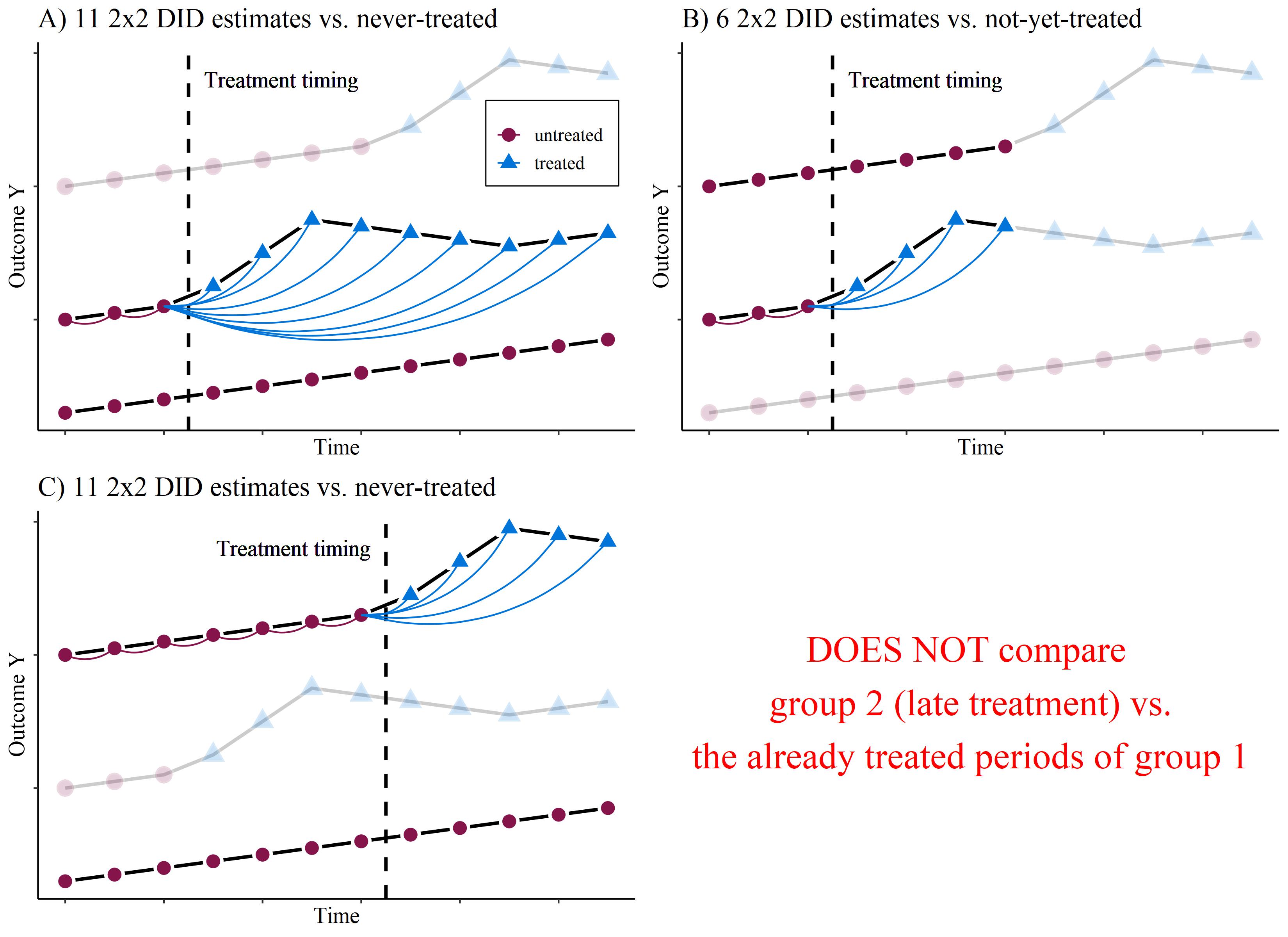}
  \caption{The three contrasting pairs with 28 single $2\times 2$ DiD estimates of group- and time-specific treatment effects $\delta_{g,t}$ used in \citet{Callaway.2020}. The blue curved lines indicate the periods of each $2\times 2$ DiD estimate \textit{after} treatment, which contribute to the summary ATT measure. The red curved lines indicate \textit{pre-treatment} periods. These do not contribute to the ATT summary but reveal pre-treatment differences between treated and control groups.}
  \label{fig:Callaway}
\end{figure}

Figure~\ref{fig:Callaway} illustrates examples of possible $2 \times 2$ DiD contrasts. The curved blue lines represent the first difference in $\delta_{g,t}$ (i.e., $[E(y_{g}^{t}) - E(y_{g}^{g-1})]$), and the difference with the corresponding control group change yields the ATT estimate $\delta_{g,t}$. Panels A and B show contrasts against the never-treated, while panel C shows a contrast against the not-yet-treated. A key feature of this estimator class is that it excludes any \emph{``forbidden''} comparisons in which already-treated units serve as the control group (i.e., panel D in Figure~\ref{fig:Bacon}).

The approach also produces group- and period-specific DiD estimates for pre-treatment periods, shown as red curved lines in Figure~\ref{fig:Callaway}. While these pre-treatment contrasts are not used in the summary ATT estimate, they are useful for visually inspecting potential violations of the parallel trends assumption.

The group- and time-specific ATTs $\delta_{g,t}$ (typically numerous) can be estimated using various methods, including conventional OLS, inverse probability weighting, or doubly robust estimators \cite{Callaway.2020}. These estimates can then be aggregated into a single summary measure. For instance, one could compute a weighted average using only contrasts with never-treated units as controls (e.g., contrasts 1 and 2 in Figure~\ref{fig:Callaway}). If the number of never-treated units is small, contrasts involving not-yet-treated units (e.g., contrast 3) can be added. Alternatively, one can construct an ``event-study'' style summary by averaging $\delta_{g,t=g+e}$ across groups for a given number of periods $e$ after treatment adoption. These averages can be weighted equally across groups and time, or weighted by the relative frequency of each group $g$ in period $t$.

\citet{Sun.2021} propose a closely related estimator. The main difference is that their approach is regression-based and resembles an event-study design. \citet{Sun.2021} use a TWFE model with interaction terms between relative treatment timing $e = t - g$ and treatment-timing groups $g$. When no never-treated group is available, they suggest using the last-treated group as a control, whereas \citet{Callaway.2020} rely on the full set of not-yet-treated groups. However, when using the never-treated as the control group, the estimators by \citet{Sun.2021} and \citet{Callaway.2020} are equivalent for balanced panels, although they differ in how they incorporate additional covariates.

Finally, \citet{DeChaisemartin.2020} propose an estimator for settings with treatment reversals, where treatment can switch on and off \cite[see also][]{Imai.2023}. In staggered adoption settings, their estimator is equivalent to that of \citet{Callaway.2020}. Since our focus is on staggered treatment adoption, we do not include this estimator in our simulations.

\subsection{Model-Based Imputation}

\citet{Borusyak.2024} propose an estimator that imputes counterfactual outcomes for each treated observation. All post-treatment unit-periods for treated units are treated as if their counterfactual outcomes (i.e., what would have occurred in the absence of treatment) are missing data to be imputed. The estimator proceeds in three steps.

First, a ``control'' TWFE model is estimated using only observations from never-treated and not-yet-treated units. This model is specified flexibly, potentially incorporating all pre-treatment or time-varying covariates that are not causally affected by treatment, as well as interactions among them. In the second step, this control model is used to impute the counterfactual outcome $\hat{Y}_{i,t}(0)$ for each treated unit $i$ at time $t$. The difference between the observed outcome and the imputed counterfactual yields the individual treatment effect: $\delta_{i,t} = Y_{i,t}(1) - \hat{Y}_{i,t}(0)$. In the third and final step, these unit- and time-specific treatment effects are aggregated into a summary estimate using a weighted average.

Formally, the model begins with a flexible TWFE specification:

\begin{equation}
\label{eqn:borusyak}
Y_{it} = A_{it}^{'}\lambda_i + X_{it}^{'}\delta + D_{it}(\Gamma_{it}^{'}\theta) + \epsilon_{it}  
\end{equation}

\noindent where $A_{it}^{'}$ includes unit indicators and potentially covariates unaffected by treatment, and $\lambda_i$ are unit-specific coefficients. Thus, $A_{it}^{'}\lambda_i$ captures unit fixed effects and possibly their interactions with other covariates. The term $X_{it}^{'}\delta$ includes time fixed effects and time-varying covariates, which must be strictly exogenous to treatment. The term $(\Gamma_{it}^{'}\theta)$ specifies the functional form of the treatment effect, allowing for heterogeneity across time or units. By default, $\Gamma = \textbf{I}_{N1}$, meaning each treated unit-time pair receives its own treatment effect (similar to $\delta_{g,t}$ in Equation~\ref{eqn:callaway1}). The algorithm proceeds as follows \cite{Borusyak.2024}:

\begin{enumerate}
\item For each treated observation, estimate the expected untreated potential outcome $\hat{Y}_{it}(0)$ using a TWFE model ($A_{it}^{'}\lambda_i + X_{it}^{'}\delta$) fit only on never-treated and not-yet-treated observations. Since these units have $D_{it} = 0$, the treatment term $D_{it}(\Gamma_{it}^{'}\theta)$ drops from Equation~\ref{eqn:borusyak}.
\item For each treated observation, compute $\hat{\delta}_{it} = Y_{it} - \hat{Y}_{it}(0)$, which represents the estimated unit-specific causal effect of treatment.
\item Estimate the overall treatment effect as a weighted sum: $\hat{\delta} = \sum_{it} w_{it} \hat{\delta}_{it}$.
\end{enumerate}

The estimator uses all available pre-treatment periods -- including those far from the treatment date -- for imputing counterfactuals. \citet{Borusyak.2024} also propose several weighting schemes $w_{it}$ for aggregating the individual effects. For example, to estimate the overall ATT, one can use uniform weights $w_{it} = 1/N_1$ across all treated units. For event-study analyses, the weights can be adjusted to estimate time-specific treatment effects $\delta_D(e)$.

\subsection{Matrix Completion}

The matrix completion approach \cite{Athey.2021} represents a class of estimators that, like model-based imputation, aims to recover counterfactual outcomes. However, it draws from the computer science literature on machine learning methods for predicting missing entries in a matrix \cite[see also][]{Xu.2017}. Again, the missing entries correspond to the potential outcomes of treated units had they not been treated. Similar to the model-based imputation method discussed above, matrix completion uses the observed entries in the outcome matrix for untreated periods to impute the missing entries for treated periods. 

We begin with the potential outcome matrices $Y(0)$ and $Y(1)$, where $Y(1)$ is observed when units are treated ($D_{it} = 1$), and $Y(0)$ is observed when units are untreated ($D_{it} = 0$):

$$
Y(0)=\left(
\begin{array}{cccc}
{\color{red} ?}   &  {\color{red} ?}  & \ldots &   \checkmark  \\
\checkmark        &  \checkmark       & \ldots &   {\color{red} ?}  \\
 \ldots           &  \ldots           & \ddots &   \ldots \\
{\color{red} ?}   &  \checkmark       & \ldots &  {\color{red} ?}  \\
\end{array}
\right)\hskip1cm  Y(1)=\left(
\begin{array}{cccc}
\checkmark   &  \checkmark & \ldots & {\color{red} ?}     \\
{\color{red} ?}        &  {\color{red} ?}        & \ldots &   \checkmark  \\
 \ldots           &  \ldots           & \ddots &   \ldots \\
\checkmark  &  {\color{red} ?}        & \ldots &  \checkmark  \\
\end{array}
\right)
$$

\noindent To estimate the causal effect of treatment, we need to impute the missing entries in $Y(0)$. Since the realised outcomes for treated units are observed, imputing $Y(1)$ is unnecessary. The task of recovering $Y(0)$ is known in the computer science literature as a matrix imputation problem \cite{Athey.2021}.

\citet{Athey.2021} distinguish between two types of data structures. In cross-sectional settings with many units and few time periods ($N \gg T$), the data matrix is ``thin''. Here, only the final periods are missing for treated units, and missing values can be imputed using horizontal regression -- regressing outcomes in the last periods on lagged outcomes $Y_{it-1}$. In contrast, time-series settings (e.g., synthetic control with a single treated unit) are ``fat'' ($T \gg N$), where many time periods are observed for each unit. In such cases, vertical regression is used -- regressing the treated unit’s outcomes on those of untreated units at the same time point \cite{Abadie.2010}.

In staggered DiD settings with many units and many time periods, the data matrix is more symmetric. Rather than relying solely on horizontal or vertical regression, matrix completion exploits stable patterns both across units and over time \cite{Athey.2021}. This is also the idea behind TWFE and more flexible interactive factor models \cite{Xu.2017}. In matrix notation, the conventional TWFE model can be written as:

\begin{equation}
\label{eq:twfe_mat}
\mathbf Y_{N\times T}= \mathbf L_{N \times T} + \mathbf \epsilon_{N \times T} = \left(
\begin{array}{ccccccc}
 \alpha_1 & 1  \\
\vdots  & \vdots   \\
\alpha_N  & 1  \\
\end{array}\right)
\left(
\begin{array}{ccccccc}
1 & \dots & 1   \\
\zeta_1  & \dots & \zeta_T  \\
\end{array}
\right) + \mathbf \epsilon_{N \times T}.
\end{equation}

Here, the matrix $\mathbf L$ captures additive unit and time effects $\alpha_i + \zeta_t$. Interactive fixed effects models generalise this by allowing for more flexible interactions between unit and time effects:

\begin{equation}
\label{eq:factor}
Y_{it} = \sum_{r=1}^{R} u_{ir}v_{tr} + \epsilon_{it},
\end{equation}

\noindent where $u_{ir}$ and $v_{tr}$ are unit- and time-specific factors, and $r = 1, \ldots, R$ is the number of factors. The conventional TWFE model corresponds to $R = 2$ (as in Equation~\ref{eq:twfe_mat}). In matrix form, this corresponds to decomposing $\mathbf L_{N \times T}$ into $\mathbf U_{N \times R} \times \mathbf V_{R \times T}$ \cite{Athey.2021, Xu.2017}. Matrix completion generalises this approach to settings with many missing values and large $N$ and $T$. Consider the following matrix of potential control outcomes:

$$
\mathbf Y_{N\times T}=\left(
\begin{array}{cccccccccc}
 {\color{red} ?} & {\color{red} ?} & {\color{red} ?} & \checkmark & {\color{red} ?}& \checkmark  & \dots  & {\color{red} ?}\\
\checkmark & {\color{red} ?} & {\color{red} ?} & {\color{red} ?} & \checkmark & {\color{red} ?}   & \dots & \checkmark  \\
{\color{red} ?}  & \checkmark & {\color{red} ?}  & {\color{red} ?} & {\color{red} ?} & {\color{red} ?} & \dots & {\color{red} ?}  \\
 {\color{red} ?} & {\color{red} ?} & \checkmark & \checkmark & \checkmark & \checkmark  & \dots  & {\color{red} ?}\\
\checkmark & \checkmark & {\color{red} ?} & {\color{red} ?} & \checkmark & {\color{red} ?}   & \dots & \checkmark  \\
{\color{red} ?}  & \checkmark & {\color{red} ?}  & {\color{red} ?} & {\color{red} ?} & {\color{red} ?} & \dots & {\color{red} ?}  \\
\vdots   &  \vdots & \vdots &\vdots   &  \vdots & \vdots &\ddots &\vdots \\
{\color{red} ?}  & {\color{red} ?} & {\color{red} ?} & {\color{red} ?}& \checkmark & {\color{red} ?}   & \dots & {\color{red} ?}\\
\end{array}
\right)
$$

To impute the missing entries, matrix completion does not aim to consistently estimate $u_{ir}$ and $v_{tr}$, but rather to impute the matrix $\mathbf Y$ as a function of the matrix $\mathbf L^{*}$ and measurement error $\epsilon$: 

\begin{equation}
\label{eq:mc}
\mathbf Y_{N\times T} = \mathbf L^{*} + \mathbf \epsilon, \quad \rm{with} \quad  \mathbb{E}[\epsilon| \mathbf L^{*}] = 0.
\end{equation}

\noindent We thus need to estimate $\mathbf L^{*}$. However, minimising the sum of squared residuals directly would simply reproduce $Y_{it}$ and yield treatment effects of zero. To avoid this, \citet{Athey.2021} propose regularisation to shrink $\mathbf L^{*}$ to a lower rank than $N \times T$. Among various regularisation techniques, they recommend nuclear norm minimisation. Additionally, they propose estimating fixed effects separately by defining $L^{*}_{it} = L_{it} + \alpha_i + \zeta_t$, so that the fixed effects are not penalised:

\begin{equation}
\label{eq:nuclear}
\arg\min_{L, \alpha, \zeta}\frac{1}{|\cal{O}|}
\sum_{(i,t) \in \cal{O}} \left(Y_{it} - 
L_{it} - \alpha_i - \zeta_t \right)^2+\lambda_L \| L\|_*,
\end{equation}

\noindent where $\cal{O}$ denotes the set of observed untreated entries (i.e., those with $D_{it} = 0$), $\|\cdot\|_*$ is the nuclear norm, and $\lambda_L$ is a regularisation parameter selected via cross-validation.

\subsection{Extended Two-Way Fixed Effects}

\citet{Wooldridge.2021} argues that the issues identified with TWFE are not inherent flaws of the estimator itself, but rather stem from a misspecification of the functional form. Specifically, if treatment effects vary across units or over time, the TWFE model should be specified in a way that captures this heterogeneity. To address this, \citet{Wooldridge.2021} proposes an Extended TWFE (ETWFE) specification. ETWFE incorporates all relevant interactions between the treatment indicator $D_{it}$ and time fixed effects, treatment cohorts, and potentially other covariates.

A treatment cohort is defined similarly to \citet{Callaway.2020}, where a cohort is a group of units that first receive treatment at time $t = g$. For each such group, ETWFE estimates separate treatment effects, time fixed effects, and covariate effects by including the appropriate interaction terms in the TWFE framework. Formally, the ETWFE model is specified as:

\begin{equation}
\label{eq:etwfe}
y_{it} = \eta + \zeta_t + \bm X_i(\beta + \nu_t) + \sum_{j=g} \left( \alpha_g + \bm X_i \beta_g + \mathbf{1}\{t > g\}(\alpha_g \times \zeta_t + \Bar{\bm X_g}\beta_{gt})  \right)  + \epsilon_{it}.
\end{equation}

\noindent Here, $\eta$ is the global intercept, $\zeta_t$ are time fixed effects, and $\bm X_i$ is the vector of pre-treatment covariates. The coefficients $\beta$ and $\nu_t$ capture the baseline and time-varying effects of these covariates. $\alpha_g$ denotes fixed effects for treatment cohorts, where each group $g$ receives treatment at time $g$. The indicator $\mathbf{1}\{t > g\}$ equals 1 if unit $i$ is observed after treatment and 0 otherwise. $\Bar{\bm X_g}$ represents group-mean-centered pre-treatment covariates, and $\beta_{gt}$ captures the interaction between group indicators and covariates. The term $\alpha_g \times \zeta_t$ represents interactions between cohort and time indicators, which are used to recover group- and time-specific treatment effects $\delta_{g,t}$.

ETWFE essentially saturates the standard TWFE model with all relevant interaction terms involving treatment cohorts. As a result, the individual coefficients are not directly interpretable. However, they can be used to compute \emph{marginal effects}, which summarise treatment effects averaged across all treatment cohorts, typically holding covariates at their means. This also allows for the estimation of event-study-type effects $\delta_D(e)$, where the average treatment effect is computed for $e$ periods after treatment.

\section{Monte Carlo Simulation}

The previous sections have shown that single-number summary measures -- estimated from a specification with a single treatment indicator -- from TWFE yield biased estimates when treatment effects vary over time. While the magnitude of this bias may differ, it consistently appears across both macroeconomic (relatively small $N$ and large $T$) and typical sociological panel data (with large $N$ and small $T$) contexts.

To further investigate the extent of this bias, we simulate several scenarios commonly encountered in applied research including various forms of treatment effects, effect heterogeneity and time horizon. In our main results, we examine three distinct types of dynamic treatment effects: (1) a step-level shift, (2) an inverted-U shape, and (3) a trend-breaking effect. These patterns reflect a range of empirically observed treatment dynamics. In addition to these three main forms, we also explore alternative functional forms -- such as fading-out or gradual fading-in effects -- which yield qualitatively similar results (see Supplement \ref{suppl:fade-out}). We also compare how the novel DiD estimators perform relative to the conventional TWFE under these conditions, focusing both on single summary measures and event-time specifications that estimate treatment effects relative to the timing of treatment.

In a second set of simulations, we introduce violations of key identifying assumptions. Such violations are common in applied research, making it essential to assess the behaviours of different estimators under these conditions. The first assumption is \emph{parallel trends}, which posits that in the absence of treatment, treated and untreated units would follow similar outcome trajectories over time. The second is the \emph{no anticipation} assumption, which requires that units do not respond to treatment before it is implemented. Since these assumptions are often difficult to verify in practice, understanding how estimators behave when they are violated is of critical importance.

Table~\ref{tab:scenarios} summarises the simulation scenarios we consider. Set-up 1 serves as a baseline with a homogeneous, time-constant step-level treatment effect -- under which all estimators should perform well. Set-ups 2 and 3 introduce time heterogeneity through trend-breaking effects, with Set-up 3 adding group-specific variation in treatment intensity. Set-up 4 replaces the trend-breaking effect with an inverted-U shape. Finally, Set-ups 5 and 6 introduce violations of the identifying assumptions: anticipation in Set-up 5 and non-parallel trends in Set-up 6.

%We focus on a subset of scenarios that are particularly informative for understanding the limitations of TWFE and the strengths of its alternatives. The central issue is time heterogeneity, which is known to bias TWFE but should be addressed by more flexible estimators. Therefore, we present simulations for the three main types of time heterogeneity (step-level, inverted-U, and trend-breaking), and then introduce additional violations -- such as anticipation, omitted trending variables, and non-parallel trends -- while holding the inverted-U treatment effect constant, as this pattern is common in many social science applications.

The scenarios in Table~\ref{tab:scenarios} provide a representative and informative overview. Additional results are available in the supplementary material. Researchers interested in exploring further variations can use our replication package to generate additional scenarios.

\begin{table}
\caption{Various scenarios implemented in the simulations.}
\label{tab:scenarios}
\resizebox{\linewidth}{!}{%
\begin{tabular}{cccccc}\toprule
Set-up 	 & effect over time   & effect structure	& group-specific 	& anticipation  & parallel trends \\\midrule 
      1  & homogeneous        & step-level shift  	& no 				& no 		   	& yes \\
      2  & heterogeneous      & trend breaking  	& no				& no 		   	& yes \\
      3  & heterogeneous      & trend breaking  	& late = 0.5 early 	& no 			& yes \\
      4  & heterogeneous      & inverted-U shaped 	& late = 0.5 early 	& no	 		& yes \\
      5  & heterogeneous      & inverted-U shaped 	& late = 0.5 early 	& negative 		& yes \\ 
      6  & heterogeneous      & inverted-U shaped 	& late = 0.5 early 	& no 			& no
      \\\bottomrule
\end{tabular}
}
\end{table}

\subsection{Data Generating Process}

We employ a Monte Carlo simulation to compare the performance of the estimators across the scenarios outlined above. All simulations in the following section are based on simulations with $N = 2,000$ units observed over $T = 15$ time periods (see Supplement \ref{fig:fig_att_larget} for additional results with small $N$ and large $T$). Each scenario is run with $R = 1,000$ simulation replications, using a fixed random seed across all set-ups to ensure comparability. The data generating process is specified as follows:

%\begin{equation} \label{eq:dgp1}
%y_{it} = \alpha_i + \theta t + D_i\rho t + \sum_{g \neq 0}\mathbf{1}\{G_{it} = g\}\beta_g + \sum_{g \neq 0}\mathbf{1}\{G_{it} = g\}\gamma_g L_i + Z_{it}\beta_z + \varepsilon_{it}
%\end{equation}
%
\begin{equation} \label{eq:dgp1}
y_{it} = \alpha_i + \theta t + \rho t D_i + \sum_{g \neq 0}\mathbf{1}\{G_{it} = g\}\beta_g + \sum_{g \neq 0}\mathbf{1}\{G_{it} = g\}\gamma_g L_i + \varepsilon_{it}
\end{equation}

\noindent Here, $\varepsilon_{it} \sim \mathcal{N}(0, 0.2^{2})$ is a normally distributed idiosyncratic error term, and $\alpha_i \sim \mathcal{N}(0, 1^{2})$ are individual fixed effects. $\theta t$ and $\rho t$ represent time a general time trend and a treatment-groups specific time trend, respectively; for simplicity, we use a linear time trend. The binary indicator $D_i \in \{0, 1\}$ denotes whether an individual is ever treated ($D_i = 1$) or never treated ($D_i = 0$), and the indicator $L_i \in \{0, 1\}$ equals one if an individual is in the late-treated group. Both are constant over time. The interaction term $\rho t D_i$ introduces a non-parallel trend for treated units, independent of treatment timing. $G_{it}$ denotes the time relative to treatment, with $G_{it} = 1$ indicating the first post-treatment period, and $\mathbf{1}\{\cdot\}$ is the indicator function. The interaction $\sum_{g \neq 0}\mathbf{1}\{G_{it} = g\}\gamma_g L_i$ introduces group-specific treatment effects. 
%$Z_{it}$ represents additional covariates, which are omitted in the main scenarios (see below).

Treatment assignment and timing are determined in a two-step process. First, the overall treatment indicator $D_i$ is generated for each unit $i = 1, \ldots, N$ as a logistic function of the individual fixed effects $\alpha_i$:

\begin{equation} \label{eq:dgp2}
D_{i} = \mathrm{Bernoulli} \left( \frac{1}{1 + \exp^{- \lambda \alpha_i}}\right),
\end{equation}

\noindent where $\lambda$ is a scaling parameter, set to $\lambda = 5$ in all simulations. This setup results in approximately 50\% of units being treated ($D_i = 1$) and 50\% never treated ($D_i = 0$). Since $D_i$ depends on $\alpha_i$, this introduces cross-sectional selection, requiring a DiD-type design for unbiased estimation.

Next, for each treated unit ($D_i = 1$), the timing of first treatment is determined by:

\begin{equation} \label{eq:dgp3}
d_{it} = \mathrm{Bernoulli} \left( \frac{1}{1 + \exp^{- \lambda \phi_t }} \right),
\end{equation}
%\begin{equation} \label{eq:dgp3}
%d_{it} = \mathrm{Bernoulli} \left( \frac{1}{1 + \exp^{- \lambda ( \phi_t +  Z_{it}\beta_z)}} \right),
%\end{equation}

\noindent where $\phi_t$ follows a cumulative normal distribution over time with a mean of 8 and a standard deviation of 2, centering and distributing treatment timing normally around the middle of the observation window. 
%Additional covariates $Z_{it}$ can be included if $\beta_z > 0$ in Equations~\ref{eq:dgp1} and \ref{eq:dgp3}. However, in our main simulations, we set $\beta_z = 0$ and omit covariates, as the scenario with non-parallel trends already captures the relevant dynamics.

\subsection{Results}

\begin{figure}[t]
    \centering
    \includegraphics[width=\textwidth]{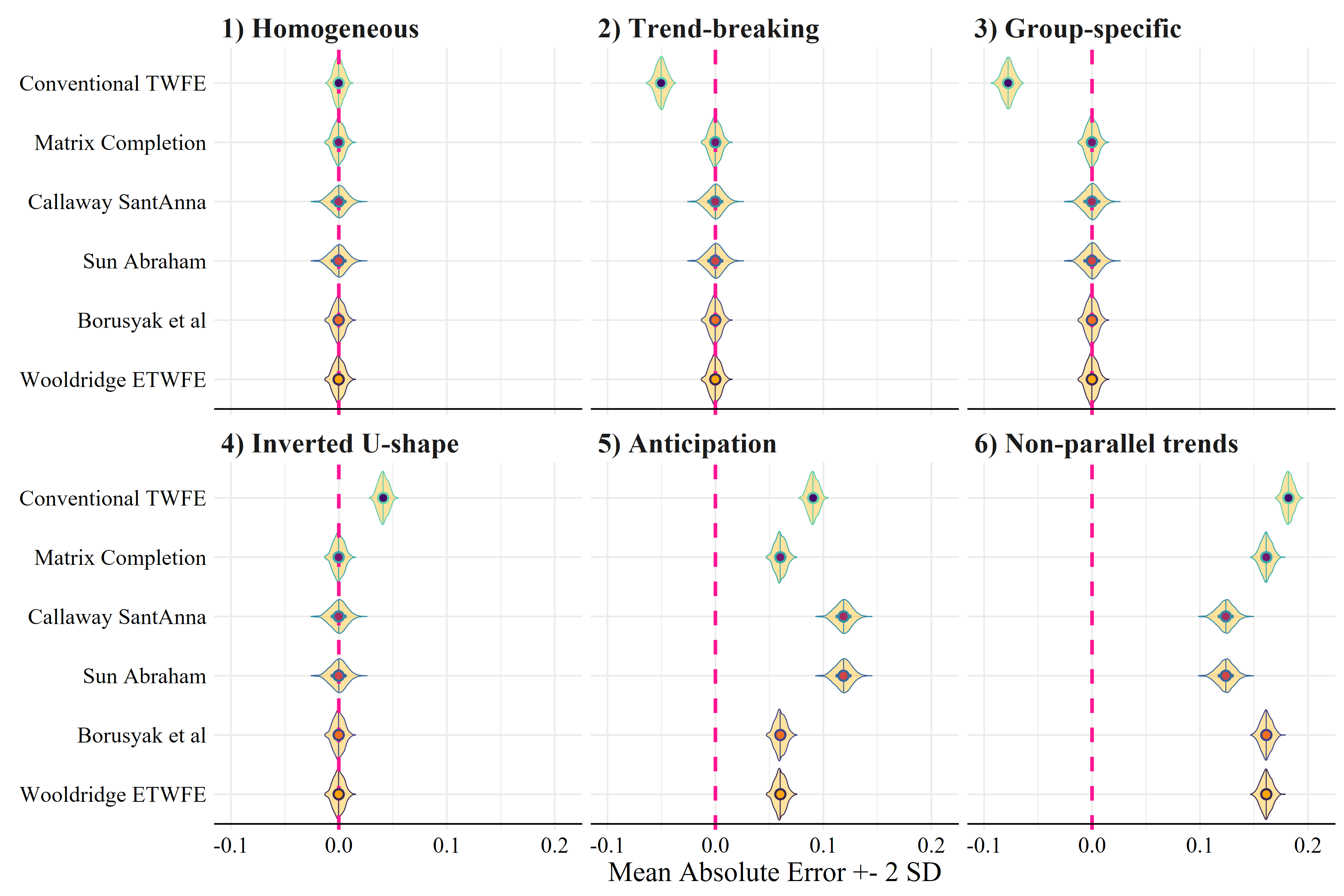}
    \caption{Monte Carlo results for single summary measures of the effect estimates. Set-ups 1 to 6 correspond to the scenarios described in Table~\ref{tab:scenarios}. Shown are the average coefficient estimates +/- two standard deviations. The yellow violin plots depict the distribution of the 1,000 individual estimates. The pink dashed line indicates the true effect according to the DGP, centered around zero.}
    \label{fig:fig_att}
\end{figure}

In the following, we present two types of results. The \emph{first type} focuses on single-number summary estimates of the treatment effect (Figure~\ref{fig:fig_att}). These summary measures are often a primary interest in applied research. For the conventional TWFE estimator, this corresponds to a model with a single binary treatment indicator (treated vs. untreated), rather than a flexible event-time specification \cite{Ludwig.2021}. When treatment effects are heterogeneous, this functional form is misspecified, leading to biased estimates -- as demonstrated earlier. In contrast, the novel dynamic DiD estimators compute frequency-weighted averages of group- and time-specific treatment effects. These estimators are designed to accommodate treatment effect heterogeneity and are therefore expected to perform better when heterogeneity is a concern. Figure~\ref{fig:fig_att} displays the mean absolute error of the single summary estimates, along with their standard deviations.

The \emph{second type} of results, shown in Figures~\ref{fig:fig2} to \ref{fig:fig5}, is based on more flexible specifications that do not rely on a single summary estimate. These models estimate time-specific treatment effects. For TWFE, this corresponds to an event-time specification, including interactions between time around the treatment and treatment group indicators. We use the first overall period and the period immediately before treatment as reference categories. The panels in Figures~\ref{fig:fig2} to \ref{fig:fig5} plot the estimated event-time coefficients along with their standard errors.

\subsubsection{Homogeneous Treatment Effect}

Set-up 1 serves as our baseline scenario: there is no anticipation, the parallel trends assumption holds, and the treatment effect is constant over time (a step-level shift). As expected, Figure~\ref{fig:fig_att} shows that the TWFE estimator is perfectly unbiased in this simple setting. The alternative estimators also perform well on average. However, the estimates from Callaway \& Sant’Anna and Sun \& Abraham exhibit slightly lower efficiency, as indicated by a somewhat larger dispersion of individual estimates around the true effect.

Although unnecessary in this scenario, applying an event-time specification with dynamic treatment interactions (see supplementary Figure~\ref{fig:fig1}) would still yield consistent results across all estimators. This is unsurprising, as Set-up 1 satisfies the identifying assumptions required by each of the methods considered.

\subsubsection{Trend-Breaking Treatment Effect}

In Set-up 2, the treatment induces a trend break -- causing a continuous increase in the outcome from the point of treatment onward. When using a single summary measure (i.e., a model with a single binary treatment indicator), the TWFE estimator produces biased results. As shown in Figure~\ref{fig:fig_att}, this results in a 28\% bias relative to the empirical ATT of 0.177. As discussed earlier, this bias arises because TWFE includes the ``forbidden'' comparison with already-treated units, thereby contaminating the control group with ongoing treatment effects.

In contrast, all of the novel DiD estimators -- explicitly designed to handle treatment effect heterogeneity -- perform well, yielding estimates with negligible average bias. These estimators accurately capture the dynamic nature of the treatment effect in the form of a trend break. Among them, the imputation-based method \cite{Borusyak.2024}, the matrix completion approach \cite{Athey.2021}, and the extended TWFE model \cite{Wooldridge.2021} demonstrate slightly higher efficiency, as indicated by narrower distributions of the individual estimates across the 1,000 replication runs.

\begin{figure}[t]
    \includegraphics[width=\textwidth]{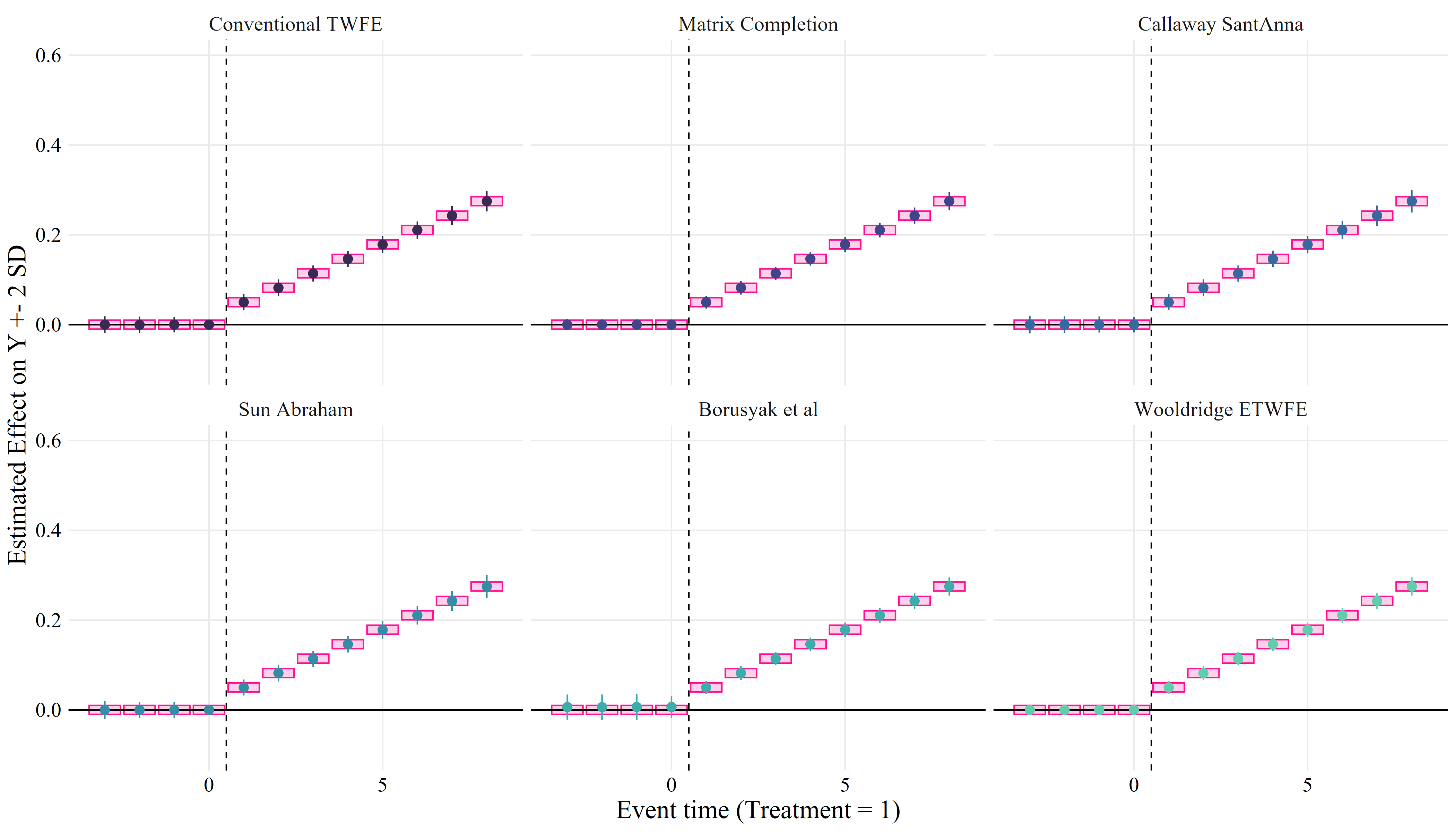} 
    \caption{Set-up 2 Monte Carlo results: parallel trends, trend-breaking treatment effects, no anticipation. Shown are the average coefficient estimates +/- two standard deviations. Pink rectangles indicate the true effect according to the DGP.}
    \label{fig:fig2}
\end{figure}

The results from the event-time specification further underscore a key point: the bias in the TWFE estimator under this scenario is due solely to misspecification of the functional form. When TWFE is specified using event-time indicators, it performs just as well as the novel dynamic DiD estimators in estimating the time-specific treatment effects (Figure~\ref{fig:fig2}). This highlights the argument made by \citet{Wooldridge.2021} that TWFE can still be valid when treatment heterogeneity is properly modelled.

\subsubsection{Trend-Breaking and Group-Specific Treatment Effect}

In Set-up 3, we extend the previous scenario by introducing group-specific treatment effects in addition to the time-heterogeneous, trend-breaking treatment pattern. As shown in Figure~\ref{fig:fig_att}, this added layer of heterogeneity induces further bias in the TWFE estimator when using a single-number summary measure (now equal to 37\% bias). Specifically, the TWFE estimate is now affected not only by the time dynamics of the treatment effect but also by the unequal magnitude of effects across treatment cohorts. In contrast, all alternative DiD estimators remain unaffected by the additional layer of heterogeneity.

The event-time specification results (Figure~\ref{fig:fig3}) appear broadly similar to those in the previous set-up. However, a subtle difference emerges: the TWFE estimates for later post-treatment periods -- particularly in the final year shown (eight years after treatment) -- are slightly biased downward. This occurs because we have specified that early-treated units experience a stronger treatment effect than late-treated units. Given the finite time horizon of 15 periods, we observe more early-treated units (with stronger effects) in later post-treatment periods than late-treated units (with weaker effects). While the dynamic DiD estimators correctly weight these observations according to their actual group-time-specific frequencies, the TWFE estimator underweights the early-treated group in these later periods. This misweighting results in a slight downward bias. Note that this bias becomes stronger when we define only two single treatment timings instead of distributing the treatment across the observation window (see Supplement \ref{suppl:twotime}). In this case, selection and misweighting becomes more severe, as we cannot observe any late-treated (treated in period 12 out of 15) from five years after treatment onwards. Such group specific treatment effects already cause a bias in TWFE when treatment-effects are not dynamic but follow a step-level function such as in set-up 1 (see supplementary results, Figure~\ref{fig:group-specific}). 

It is worth noting, however, that the magnitude of this bias remains relatively small -- even though we specified the treatment effect for early-treated units to be twice as large as that for late-treated units. This suggests that TWFE may still perform reasonably well in the presence of moderate group-level heterogeneity, though it remains inferior to the more flexible alternatives.

\subsubsection{Inverted-U Shaped and Group-Specific Treatment Effect}

\begin{figure}[t]
    \centering
    \includegraphics[width=\textwidth]{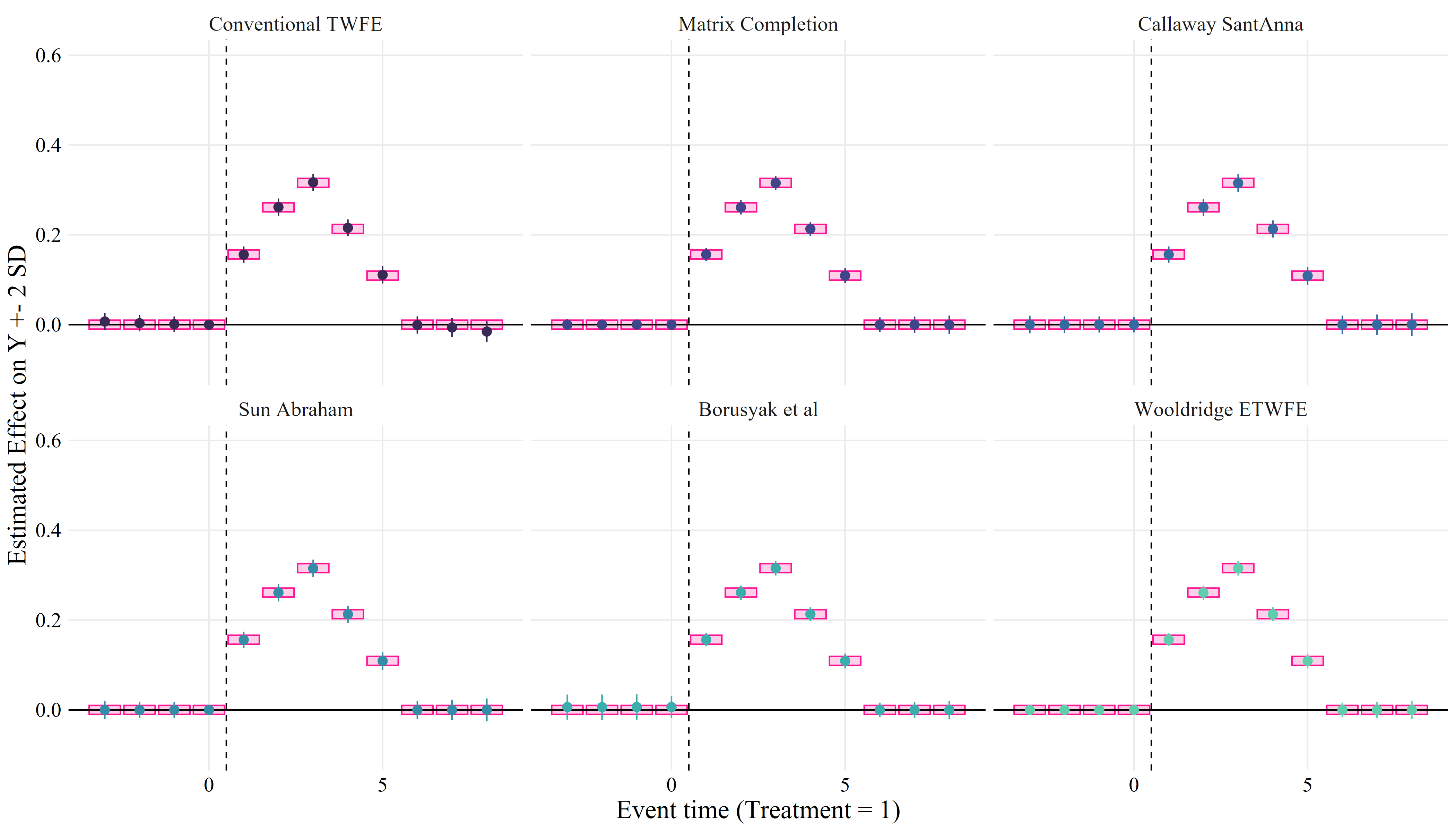}
    \caption{Set-up 4 Monte Carlo results: group-specific and time-heterogeneous (inverted-U) treatment effects with parallel trends and no anticipation. Shown are the average coefficient estimates +/- two standard deviations. Pink rectangles mark the true effect according to the DGP.}
    \label{fig:fig4} 
\end{figure}

As discussed earlier, trend-breaking treatment effects are relatively rare. In many social science applications, treatment effects tend to increase initially and then gradually fade out. Set-up 4 captures this more realistic dynamic by introducing time heterogeneity in the form of an inverted-U shaped treatment effect (Figure~\ref{fig:fig4}).

When relying on a single summary measure (Figure~\ref{fig:fig_att}, panel 4), the TWFE estimator again produces biased results, with an average relative bias of 32\% (an absolute bias of 0.041 given an empirical ATT of 0.128). The source of this bias is the same as in previous scenarios: the inclusion of ``forbidden'' comparisons and the inability of a single treatment indicator to capture time-varying effects. However, due to the inverted-U shape -- where treatment effects decline after the third year -- the direction of the bias is now positive rather than negative. This highlights an important property: the sign of the bias in the TWFE summary estimate depends critically on the specific pattern of treatment effect heterogeneity over time (see supplementary Figure~\ref{fig:fig_att_fadeout} for an alternative fading-out pattern). All other estimators -- Matrix Completion, Callaway \& Sant’Anna, Sun \& Abraham, Borusyak et al., and ETWFE -- remain unaffected by the shape of the treatment effect and perform similarly to previous scenarios, yielding unbiased estimates.

The event-time specification results (Figure~\ref{fig:fig4}) mirror the pattern observed in Set-up 3. A slight downward bias appears in the final post-treatment period for TWFE, again due to the unequal weighting of early- and late-treated units with differing treatment intensities.

\begin{figure}[t]
    \centering
    \includegraphics[width=\textwidth]{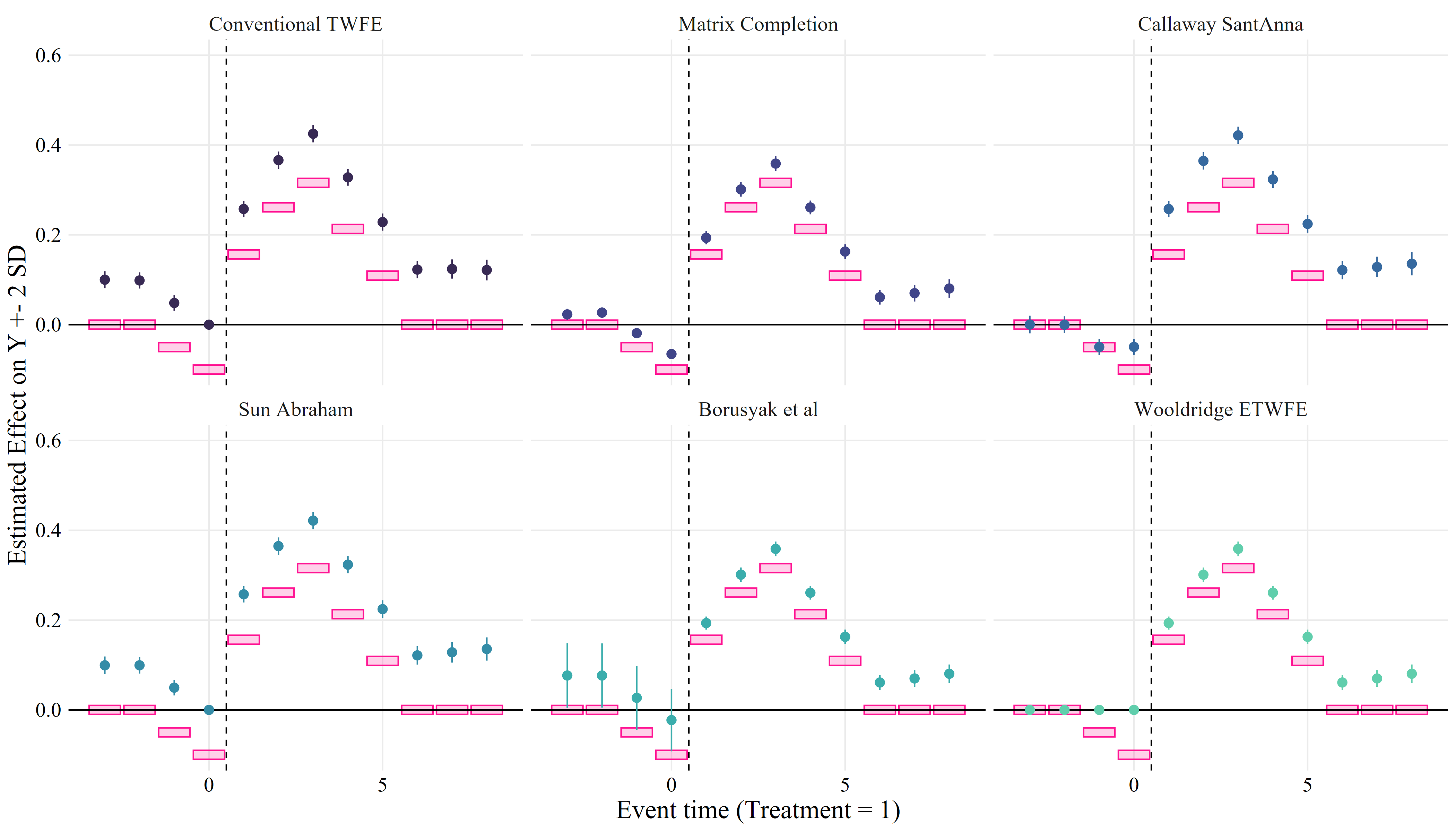}
    \caption{Set-up 5 Monte Carlo results: parallel trends, group-specific, time-heterogeneous (inverted-U) treatment effects, negative anticipation effect. Shown are the average coefficient estimates +/- two standard deviations. Pink rectangles mark the true effect according to the DGP.}
    \label{fig:fig5}
\end{figure}

\subsubsection{Anticipation}

Up to this point, we have examined scenarios involving varying degrees of treatment effect heterogeneity, while maintaining the core identifying assumptions of the novel DiD estimators -- namely, the absence of anticipation and the validity of parallel trends between treated and control units. However, these assumptions are often difficult to fully satisfy in applied research. In Set-up 5, we explicitly challenge these assumptions by introducing a (negative) anticipation effect: the outcome declines in the two periods preceding treatment.

Since all estimators rely on the no-anticipation assumption, they all produce biased estimates -- both for the single summary treatment effect (Figure~\ref{fig:fig_att}, panel 5) and for the more flexible event-time specifications (Figure~\ref{fig:fig5}).

Looking first at the single summary measures (Figure~\ref{fig:fig_att}, panel 5), the bias is most pronounced for the disaggregation-based estimators by Callaway \& Sant’Anna and Sun \& Abraham, each showing a relative bias of approximately 92\%. In contrast, Wooldridge’s ETWFE, Borusyak et al., and Matrix Completion perform considerably better under this violation, all showing a relative bias of 47\%. Interestingly, in this scenario, the conventional TWFE estimator exhibits a lower average bias (70\%) than the disaggregation-based methods.

It is important to note that all estimators could, in principle, be adapted to account for anticipation effects. For example, one could redefine the reference period to be three years before treatment, thereby avoiding the contaminated pre-treatment periods. This adjustment would mitigate the bias in both the novel DiD estimators and the conventional TWFE.

Turning to the event-time specifications (Figure~\ref{fig:fig5}), all estimators rely on at least one pre-treatment period as the reference category and implicitly assume that the treatment effect is zero in that period. If anticipation is not explicitly modelled, this leads to biased estimates of both the anticipation effect and the post-treatment effect. Among the estimators, Matrix Completion, Callaway \& Sant’Anna, and Borusyak et al. are somewhat more flexible in capturing pre-treatment dynamics. Although Wooldridge’s ETWFE performs similarly in estimating post-treatment effects, it does not estimate pre-treatment placebo effects and thus assumes them to be zero. Nevertheless, all estimators exhibit substantial bias following the onset of treatment when anticipation is present.

\subsubsection{Non-Parallel Trends}

\begin{figure}[t]
    \centering
    \includegraphics[width=\textwidth]{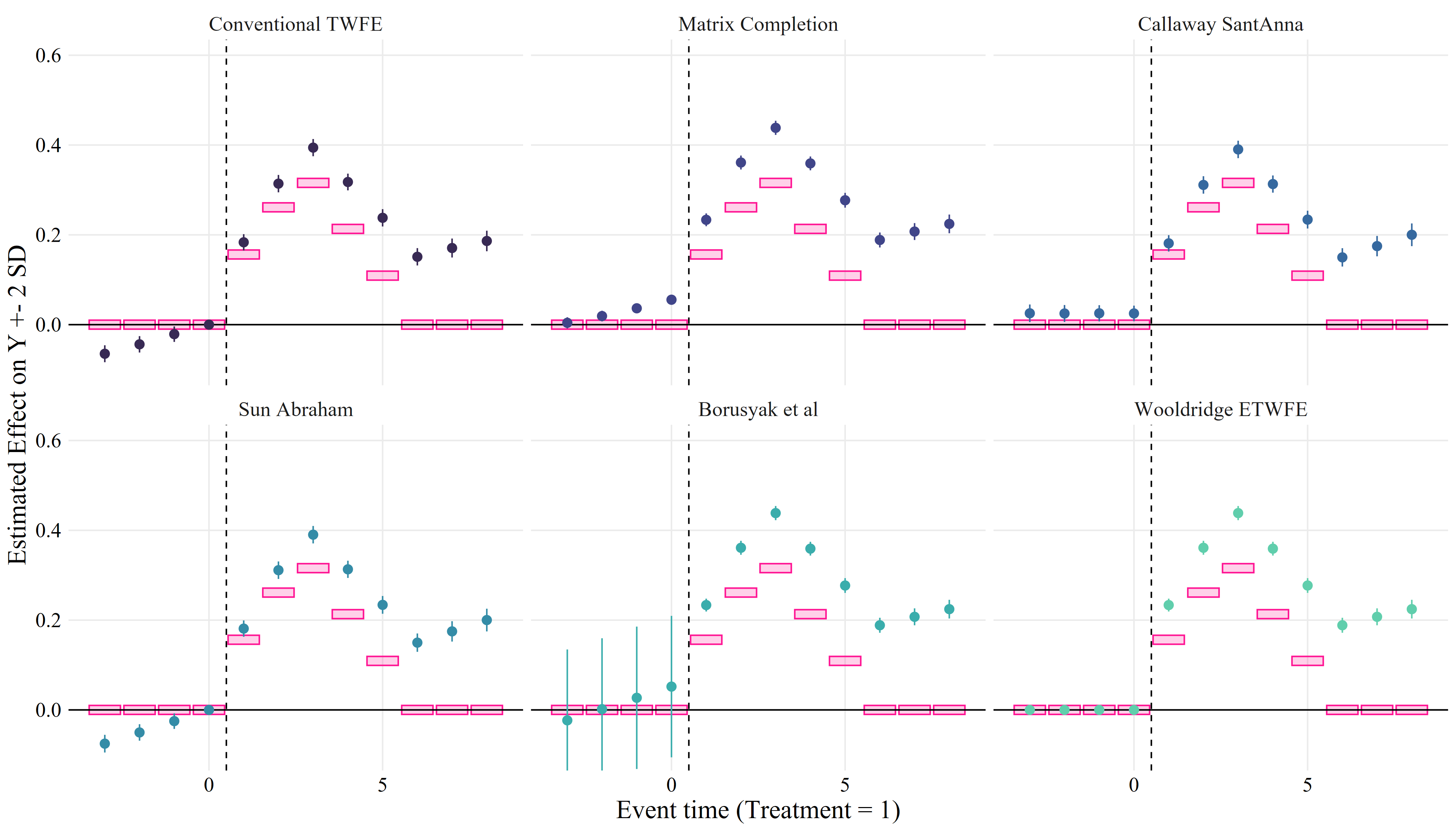}
    \caption{Set-up 6 Monte Carlo results: non-parallel trends, time-heterogeneous (inverted-U) treatment effects, no anticipation. Shown are the average coefficient estimates +/- two standard deviations. Pink rectangles mark the true effect according to the DGP.}
    \label{fig:fig6}
\end{figure}

In Figure~\ref{fig:fig6}, we remove the anticipation effects but introduce a different violation of the core assumptions: non-parallel trends. Among all identifying assumptions in DiD-like methods, the parallel trends assumption is arguably the most consequential -- and one that is frequently violated in applied research \cite{Chiu.2023}.

When this assumption is violated, all estimators exhibit substantial bias. The single summary measures in Figure~\ref{fig:fig_att} show that Sun \& Abraham and Callaway \& Sant’Anna perform slightly better than the others, with relative biases of approximately 96\%. Wooldridge’s ETWFE, Borusyak et al., and Matrix Completion all show relative biases around 126\%, only marginally better than the conventional TWFE, which reaches 142\%. While the size of the bias naturally depends on the magnitude of the trend difference, we have specified a relatively modest deviation: the treated group experiences a 12.5\% stronger time trend (e.g., increasing from 0.2 to 3.375 instead of 0.2 to 3 over 15 years). Even this moderate violation results in a bias that far exceeds those observed in previous scenarios. This highlights the centrality of the parallel-trends assumption. This becomes even more severe in a small $N$, large $T$ setting (see Figure~\ref{fig:fig_att_larget}, but be aware of the extended x-axis).

Sun \& Abraham and Callaway \& Sant’Anna are somewhat less affected by the violation of parallel trends because they rely on a single pre-treatment period as the reference category, making them less sensitive to longer pre-treatment trends. However, this also makes them more vulnerable to anticipation effects (as seen in Figure~\ref{fig:fig5}). In contrast, Wooldridge ETWFE, Borusyak et al., and Matrix Completion use all pre-treatment periods to impute post-treatment counterfactuals, making them more sensitive to violations of the parallel trends assumption but more robust to short-term anticipation effects.

Turning to the event-time specifications (Figure~\ref{fig:fig6}), all estimators again suffer from substantial bias. Although some estimators capture the pre-treatment trend differences relatively well, none are able to recover the true treatment effects accurately. Once again, Sun \& Abraham and Callaway \& Sant’Anna perform slightly better than the remaining estimators.

Importantly, the pre-treatment placebo tests should raise red flags in this scenario. With the exception of Wooldridge ETWFE, all estimators provide pre-treatment placebo tests. Matrix Completion, Sun \& Abraham, Callaway \& Sant’Anna, and the conventional TWFE all suggest violations of the parallel trends assumption through significant pre-treatment effects. Borusyak et al. is a notable exception in this regard. However, it is important to emphasise that such placebo tests are not foolproof -- especially when non-parallel trends are combined with anticipation effects. Moreover, these tests can be applied to the novel DiD estimators as well as to the conventional method. For a detailed discussion on testing the parallel trends assumption, see \citet{Roth.2022a}.

\section{Discussion}

Difference-in-differences (DiD) and the Two-Way Fixed Effects (TWFE) framework remain foundational tools for social scientists interested in causal inference. However, the application of TWFE in \emph{staggered designs} -- where treatment is implemented at different times across units -- has come under increased scrutiny in recent years \cite{Goodman-Bacon.2021, Wooldridge.2021}. A growing body of econometric research has formally demonstrated that TWFE can yield biased estimates when treatment effects vary across time and across groups. In response, several alternative estimators have been proposed to address this issue. Yet, the emergence of this new literature has also led to some confusion among applied researchers -- particularly regarding the limitations of conventional TWFE and the capabilities of the newer dynamic DiD estimators. This study aimed to clarify these issues by introducing the recent methodological developments in an accessible manner and by systematically comparing the performance of several estimators across a range of realistic, non-ideal scenarios using Monte Carlo simulations.

Our findings yield several important insights, many of which align with conclusions from recent reviews \cite[e.g.,][]{Chiu.2023, Freedman.2023, Roth.2023}. First, we emphasise that the problems associated with TWFE are not inherent to the estimator itself, but rather stem from model misspecification when treatment effects are dynamic over time \cite{Wooldridge.2021}. For instance, when treatment effects follow a trend-breaking or inverted-U pattern, using a single binary treatment indicator leads to biased estimates. The direction and magnitude of this bias depend on both the distribution of treatment timing and the functional form of the treatment effect \cite{Goodman-Bacon.2021}. 

However, when TWFE is specified using an event-time structure -- where treatment effects are modeled relative to the timing of treatment -- conventional TWFE can yield consistent estimates even under time heterogeneity. While some bias still arises in later post-treatment periods due to additional group-specific heterogeneity, this bias was rather modest across the situations that we tested here. Applied researchers should nonetheless exercise caution when interpreting long-run effects in event-time specifications, as bias tends to accumulate in periods further away from the treatment. Given the limited time horizon of most panel datasets, the groups contributing to these later-period estimates become increasingly selective, and TWFE tends to assign incorrect weights to these groups, leading to biased estimates.

We also observe meaningful differences across the newer dynamic DiD estimators. First, Sun \& Abraham and Callaway \& Sant’Anna are more sensitive to anticipation effects, as they rely on a single pre-treatment period as the reference category -- typically the period immediately before treatment. While this reference period can be adjusted with prior knowledge of anticipation, the same flexibility applies to all estimators, including TWFE. Second, Matrix Completion, Borusyak et al., and Wooldridge ETWFE are more sensitive to violations of the parallel trends assumption, as they use all pre-treatment periods to impute post-treatment counterfactuals. This design choice increases their vulnerability to diverging trends. In short, there is a trade-off: estimators that are more robust to anticipation effects tend to be more sensitive to non-parallel trends, and vice versa. When anticipation is likely, Borusyak et al., Wooldridge ETWFE, and Matrix Completion may offer more reliable estimates. Third, the disaggregation-based methods by Sun \& Abraham and Callaway \& Sant’Anna tend to be slightly less efficient (in terms of their variance) than Matrix Completion, Borusyak et al., and Wooldridge’s ETWFE across all scenarios.

\subsection*{Practical Advice for Applied Researchers}

For applied researchers working with staggered treatment designs, our results suggest a few key takeaways. First, when treatment effects are expected to vary over time, it is essential to move beyond a single binary treatment indicator. Event-time specifications -- whether implemented through TWFE or newer estimators -- offer a more flexible and accurate approach. However, even in event-time models, researchers should be cautious when interpreting long-run effects within the TWFE framework. As the time since treatment increases, event-time TWFE may produce biased estimates.
In such cases, it is advisable to rely on more robust dynamic DiD methods, such as those proposed by \citet{Borusyak.2024} or \citet{Callaway.2020} . However, even with these estimators, researchers should remain mindful of increasing selectivity in later post-treatment periods -- particularly in unbalanced panels -- where fewer units contribute to the estimates of the time-specific treatment effects.

Second, the choice of estimator should be guided by the structure of the data and the plausibility of key identifying assumptions. If anticipation effects are likely, estimators such as Borusyak et al., Wooldridge ETWFE, or Matrix Completion are preferable, as they are more robust to anticipation effects in the immediate pre-treatment periods, and also if anticipation is incorrectly modelled. Conversely, if the parallel trends assumption is in doubt, disaggregation-based methods like Callaway \& Sant’Anna or Sun \& Abraham are less sensitive to such violations, though they are not entirely immune. We encourage researcher to conduct placebo tests for pre-treatment effects, despite those tests comping with their own shortcoming \cite{Roth.2022a}. If possible, using not-yet-treated units rather than never-treated units as the control group can improve the plausibility of the parallel trends assumption. Another strategy is to explicitly model group-specific time trends, which can help account for systematic differences in outcome trajectories across treatment cohorts \cite{Ruttenauer.2023}.

Ultimately, no estimator is universally superior -- each involves trade-offs. A careful understanding of the data-generating process, combined with transparent robustness checks, remains the most reliable strategy for credible causal inference in staggered DiD designs. Our simulation results underscore that treatment effect heterogeneity can pose a serious challenge in applied research. However, addressing heterogeneity should not come at the expense of overlooking the parallel trends assumption, which remains a fundamental pillar of identification and should be a central concern for applied researchers.

%Overall, we believe that some of the recent literature about the potential problems of TWFE do not do full justice to TWFE. There are specific situations with time-heterogeneous / dynamic treatment effects in which a miss-specified TWFE fails. As we have shown the 
%
%However, checking for heterogeneous treatment effects by flexible time-varying functions, even within a conventional TWFE framework, can easily uncover this bias, and indeed may produce additional insights. Moreover, our results resonate with the conclusions of \citet{Chiu.2023} that the main threat to applied research remains the violation of the parallel trends assumption. A violation of parallel trends induces a more severe bias than miss-specifying time heterogeneity and is harder to address. Importantly, the new dynamic DiD estimators suffer from non-parallel trends as much as TWFE does. The novel dynamic DiD estimators thus do not provide a new panacea to the most important identification problem. In fact, focusing too much on the issues of heterogeneous treatment effects in TWFE and their fixes with the novel DiD estimators may distract the researcher from focusing on the bigger threats to causality. The new DiD estimators undoubtedly provide additional useful tools and are likely to add another layer of robustness of findings in applied research. However, potential violations of the important assumptions of treatment exogeneity need different solutions.

\clearpage

%\printbibliography
\bibliography{references}

\section*{Data Availability Statement}

A replication package with the simulation and analysis code is available on the author's Github repository [the link will be added].

\clearpage
\appendix

\setcounter{table}{0}
\renewcommand{\thetable}{S\arabic{table}}%
\setcounter{figure}{0}
\renewcommand{\thefigure}{S\arabic{figure}}%

\part*{\center Supplementary material} 

\section{Step function treatment} \label{suppl:step}

\begin{figure}[h!]
  \centering
  \includegraphics[width=0.8\linewidth]{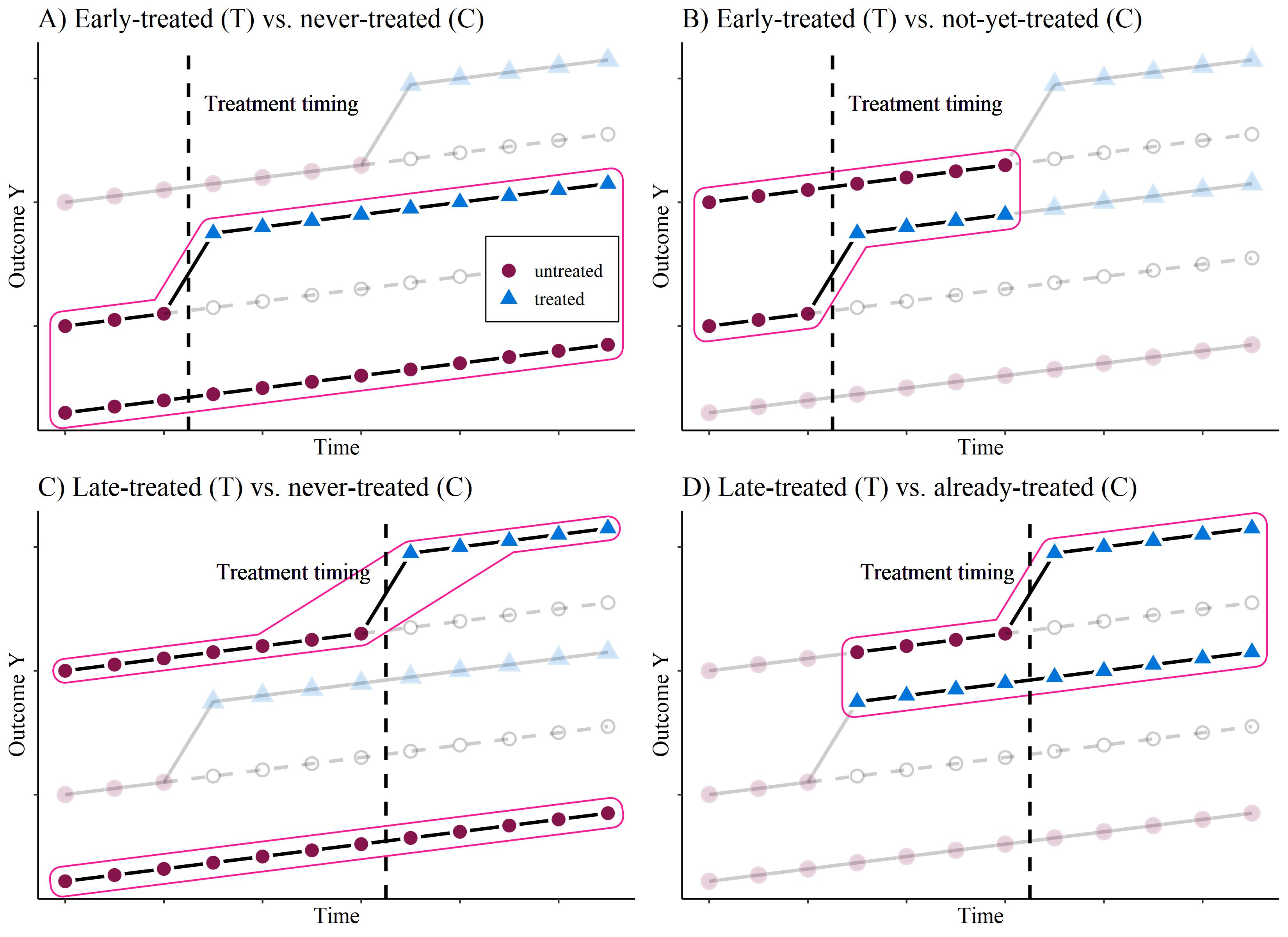}
  \caption{The four contrasts discussed by \citet{Goodman-Bacon.2021} for a non-dynamic step-function treatment.}
  \label{fig:step}
\end{figure}

\clearpage

\section{Supplementary Results} \label{suppl:desc}

\begin{figure}[h!]
    \centering
    \includegraphics[width=\textwidth]{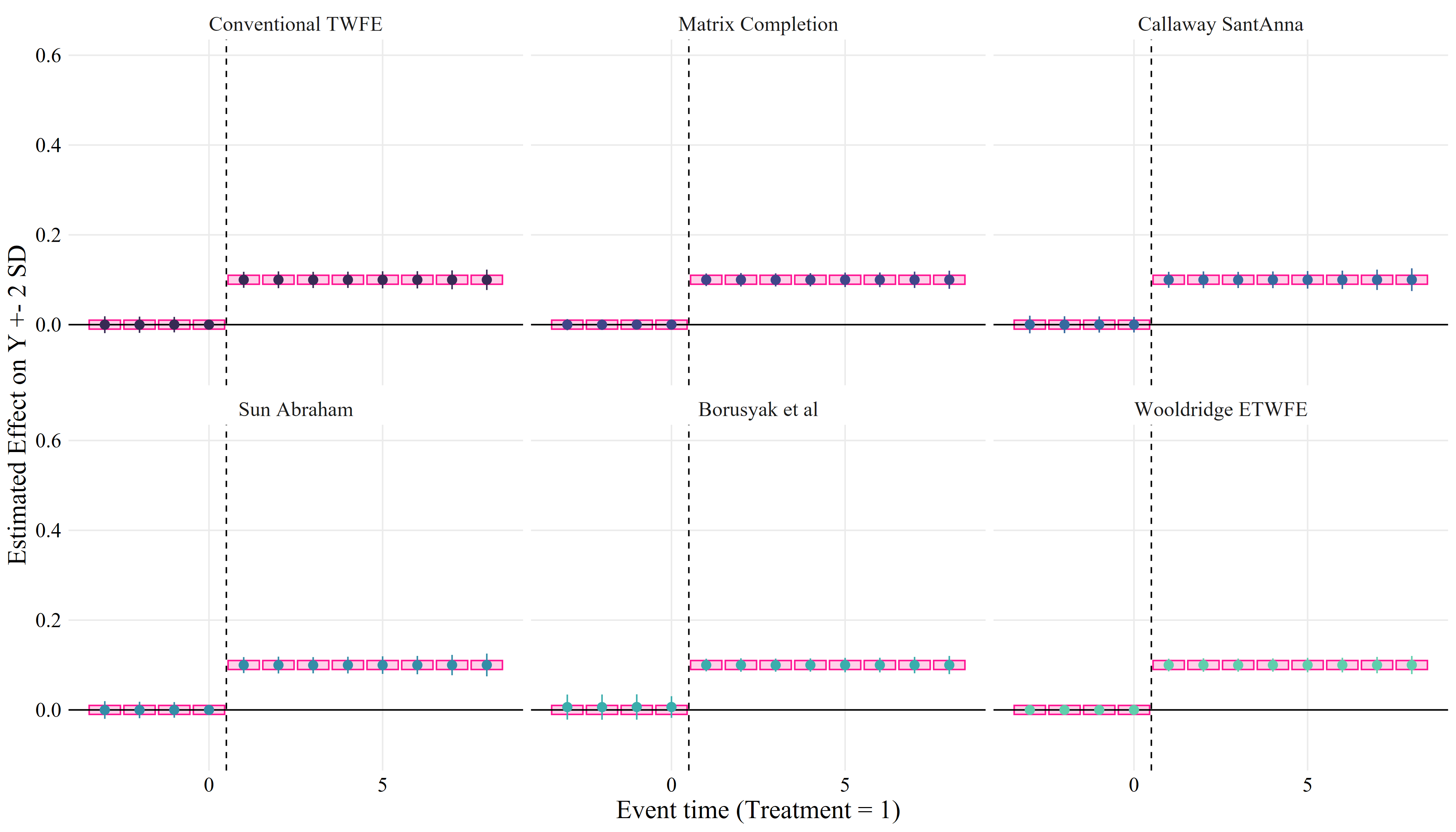} 
    \caption{Set-up 1 Monte Carlo results: parallel trends, no anticipation, homogeneous treatment effects (step-level function). Shown are the average coefficient estimates +/- two standard deviations. Pink rectangles mark the true effect according to the DGP.}
    \label{fig:fig1}
\end{figure}

\begin{figure}[t]
    \centering
    \includegraphics[width=\textwidth]{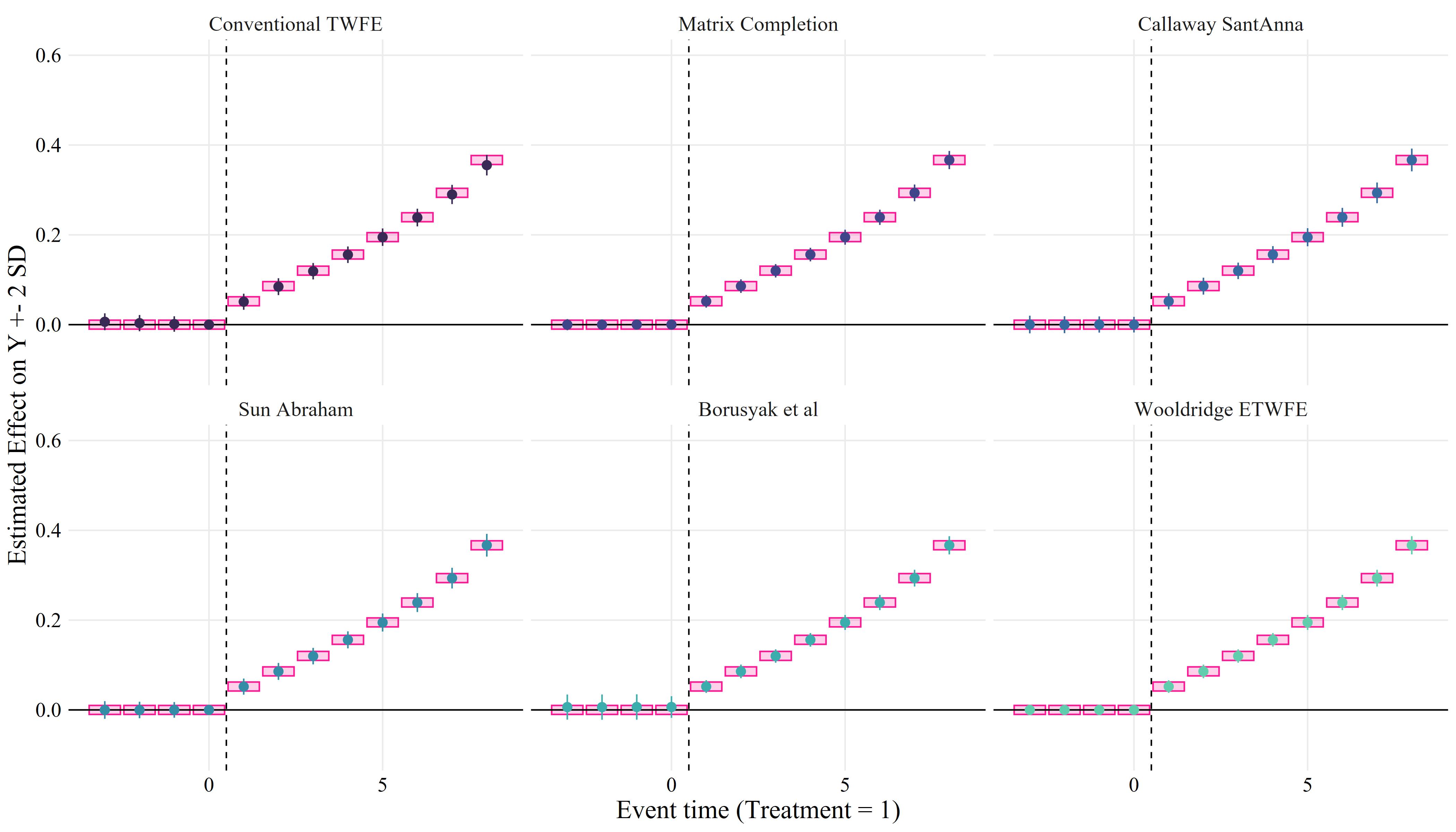}
    \caption{Set-up 3 Monte Carlo results: group-specific and time-heterogeneous (trend-breaking) treatment effects with parallel trends and no anticipation. Shown are the average coefficient estimates +/- two standard deviations. Pink rectangles mark the true effect according to the DGP.}
    \label{fig:fig3} 
\end{figure}

\begin{figure}[h!]
    \includegraphics[width=\textwidth]{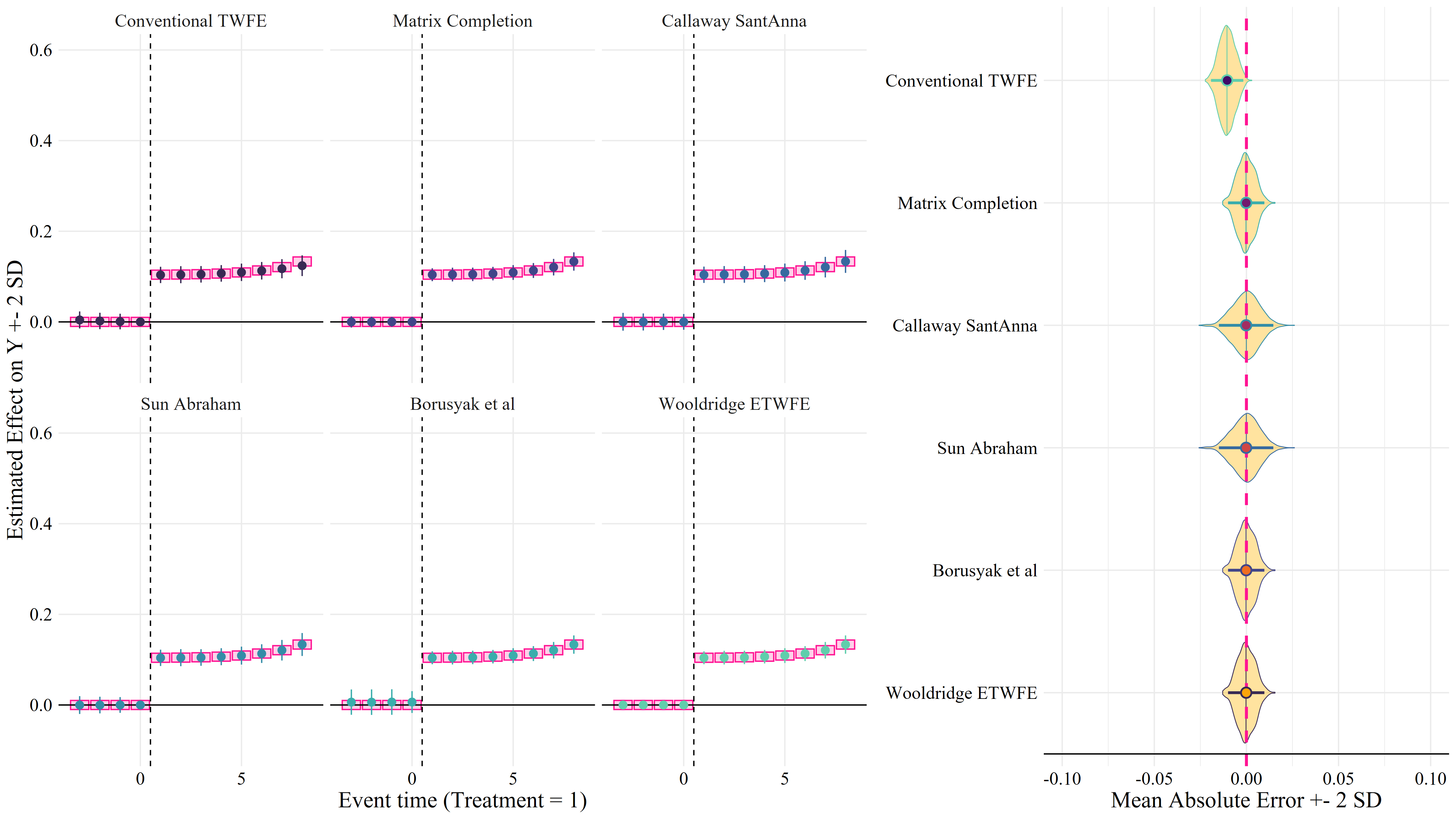} 
    \caption{Alternative Set-up 1 Monte Carlo results:  parallel trends, no anticipation, time-homogeneous treatment effects (step-level function), but group-specific treatment effects (late = 0.5 early). Shown are the average time-heterogeneous treatment effects (left) and single summary measures (right) with dispersion indicators as above. Pink rectangles / dashed line mark the true effect according to the DGP.}
    \label{fig:group-specific}
\end{figure}

%
%\begin{figure}[h!]
%    \includegraphics[width=\textwidth]{New_fade-in_combined.png} 
%    \caption{Alternative Set-up 4 Monte Carlo results: parallel trends, time-heterogeneous treatment effects (fading-in). Shown are the average time-heterogeneous treatment effects (left) and single summary measures (right) with dispersion indicators. Pink rectangles / dashed line mark the true effect according to the DGP.}
%    \label{fig:fade-in}
%\end{figure}
%
%\clearpage
%
%\begin{figure}[t]
%    \centering
%    \includegraphics[width=\textwidth]{Hettreat_FEIS_combined.png}
%    \caption{Re-runs of set-ups 3 and 6 Monte Carlo simulations: time-heterogeneous (inverse-U) treatment effects with parallel and non-parallel trends. Including estimates from Fixed Effects Individual Slopes (FEIS). Shown are the average coefficient estimates +/- two standard deviations. Pink rectangles mark the true effect according to the DGP.}
%    \label{fig:feis}
%\end{figure}

\clearpage

\section{Fade-out treatment form} \label{suppl:fade-out}

\begin{figure}[h!]
    \centering
    \includegraphics[width=\textwidth]{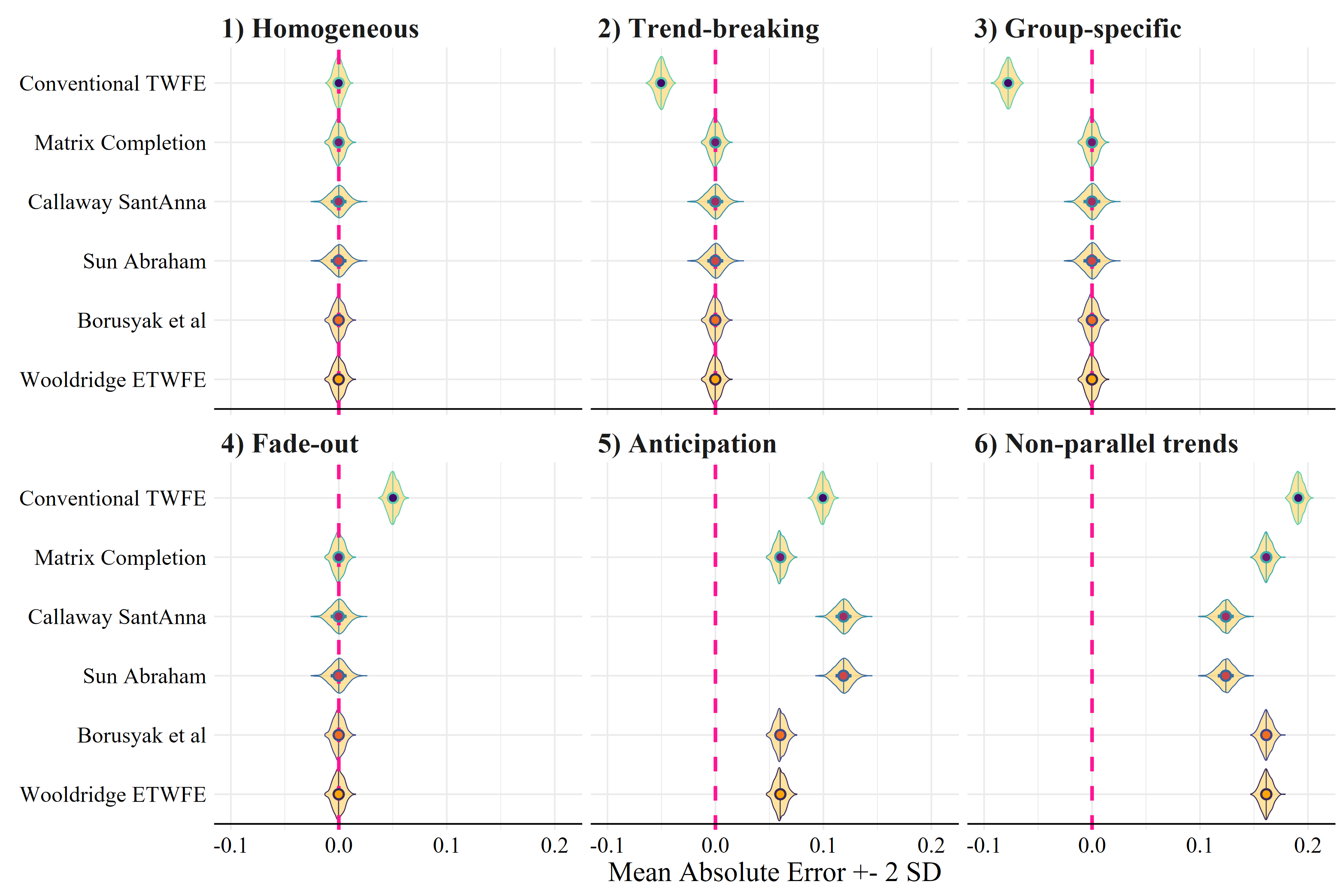}
    \caption{Monte Carlo results of single summary measure for the effect estimates. Set-ups 1 to 6 as described in Table \ref{tab:scenarios} but use a fading-out treatment form rather than an inverted u-shaped. Shown are the average coefficient estimates +/- two standard deviations. The yellow violin graphs depict the distribution of the 1,000 individual estimates. The pink dashed line marks the true effect according to the DGP, centred around zero.}
    \label{fig:fig_att_fadeout}
\end{figure}

\begin{figure}[h!]
    \includegraphics[width=\textwidth]{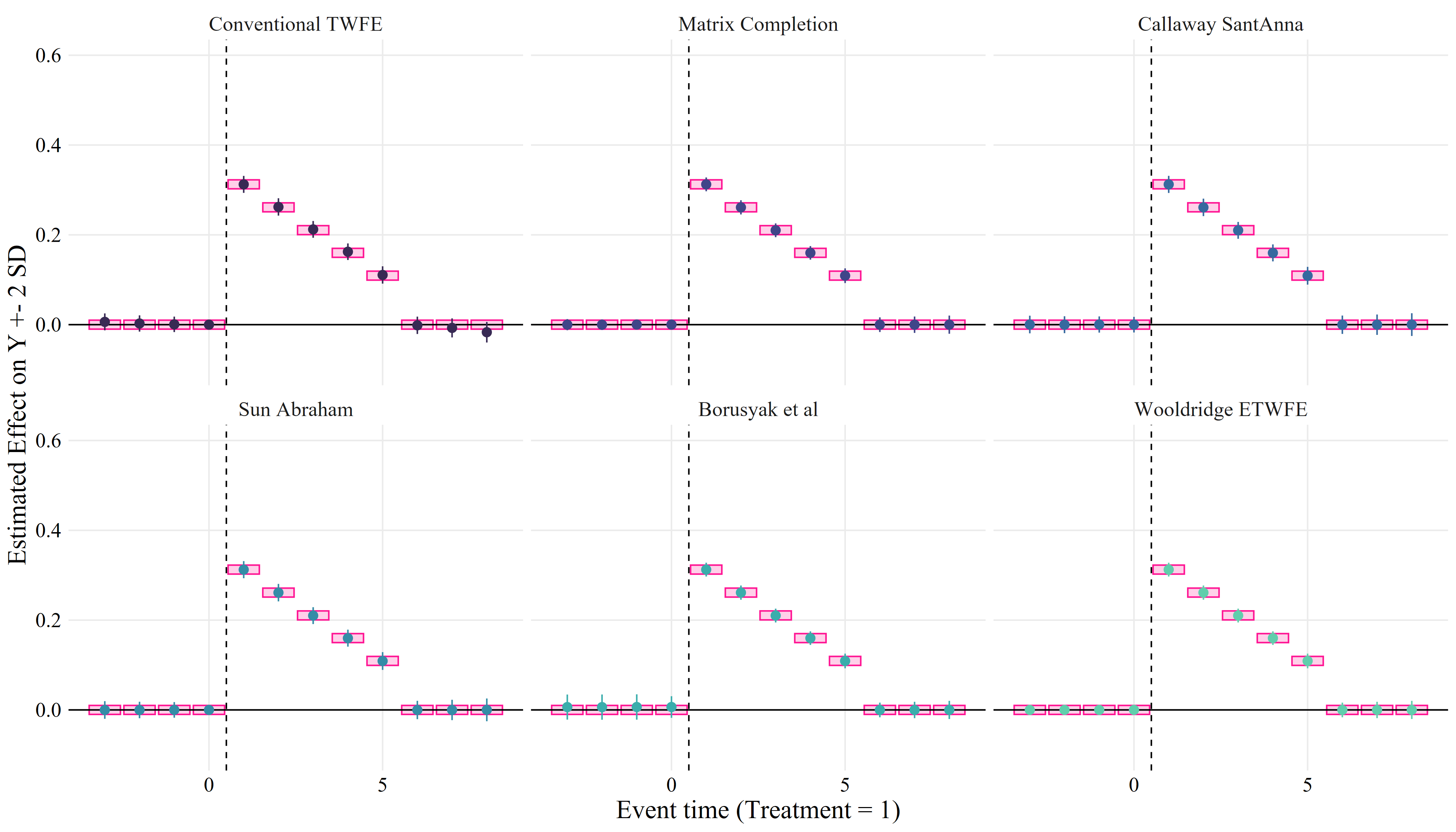} 
    \caption{Alternative Set-up 4 Monte Carlo results: parallel trends, group-specific, time-heterogeneous treatment effects (fading-out). Shown are the average coefficient estimates +/- two standard deviations. Pink rectangles mark the true effect according to the DGP.}
    \label{fig:fade-out1}
\end{figure}

\begin{figure}[h!]
    \centering
    \includegraphics[width=\textwidth]{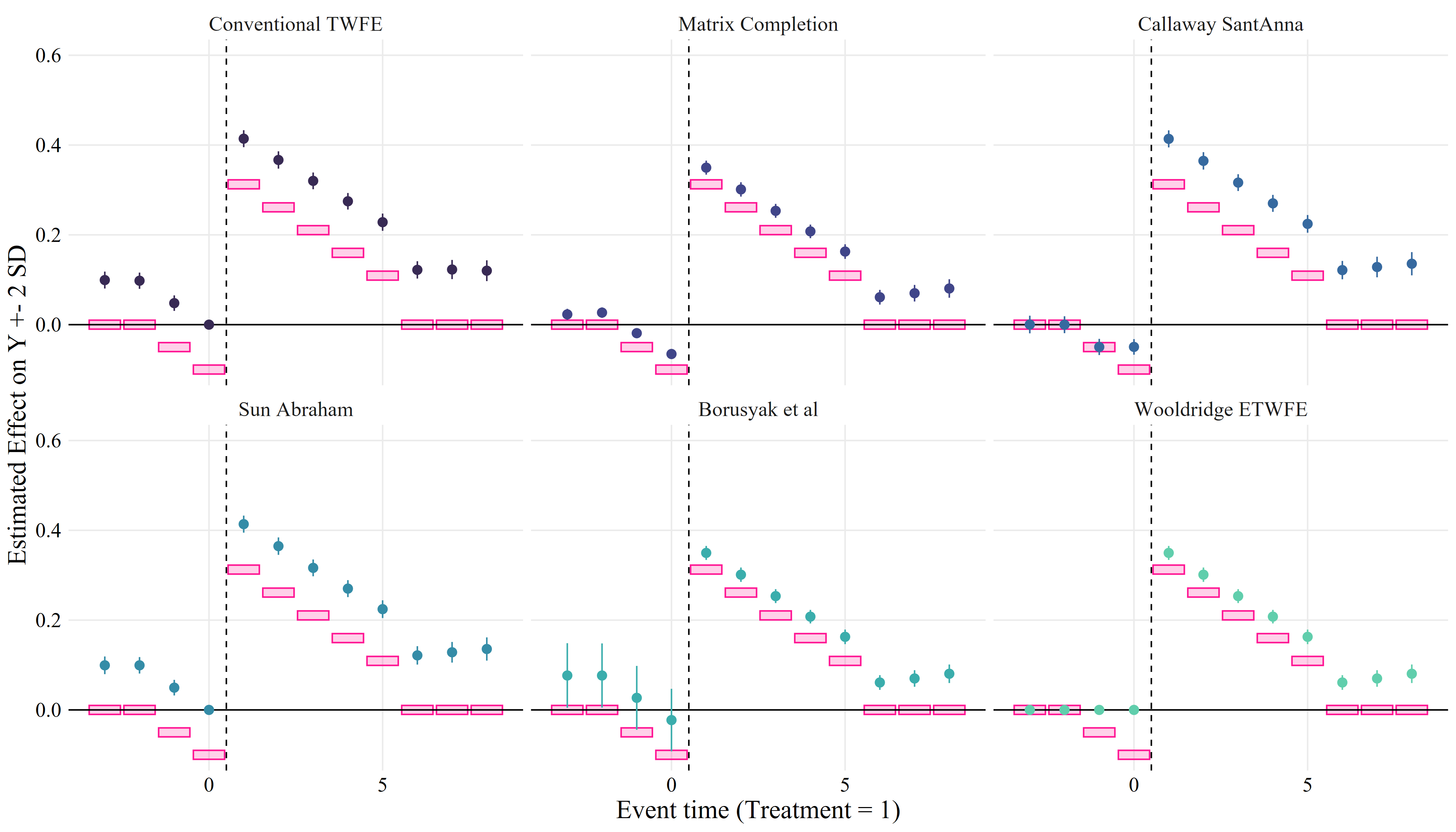}
    \caption{Alternative Set-up 5 Monte Carlo results: parallel trends, group-specific, time-heterogeneous (fading-out) treatment effects, negative anticipation effect. Shown are the average coefficient estimates +/- two standard deviations. Pink rectangles mark the true effect according to the DGP.}
    \label{fig:fade-out5}
\end{figure}

\begin{figure}[h!]
    \centering
    \includegraphics[width=\textwidth]{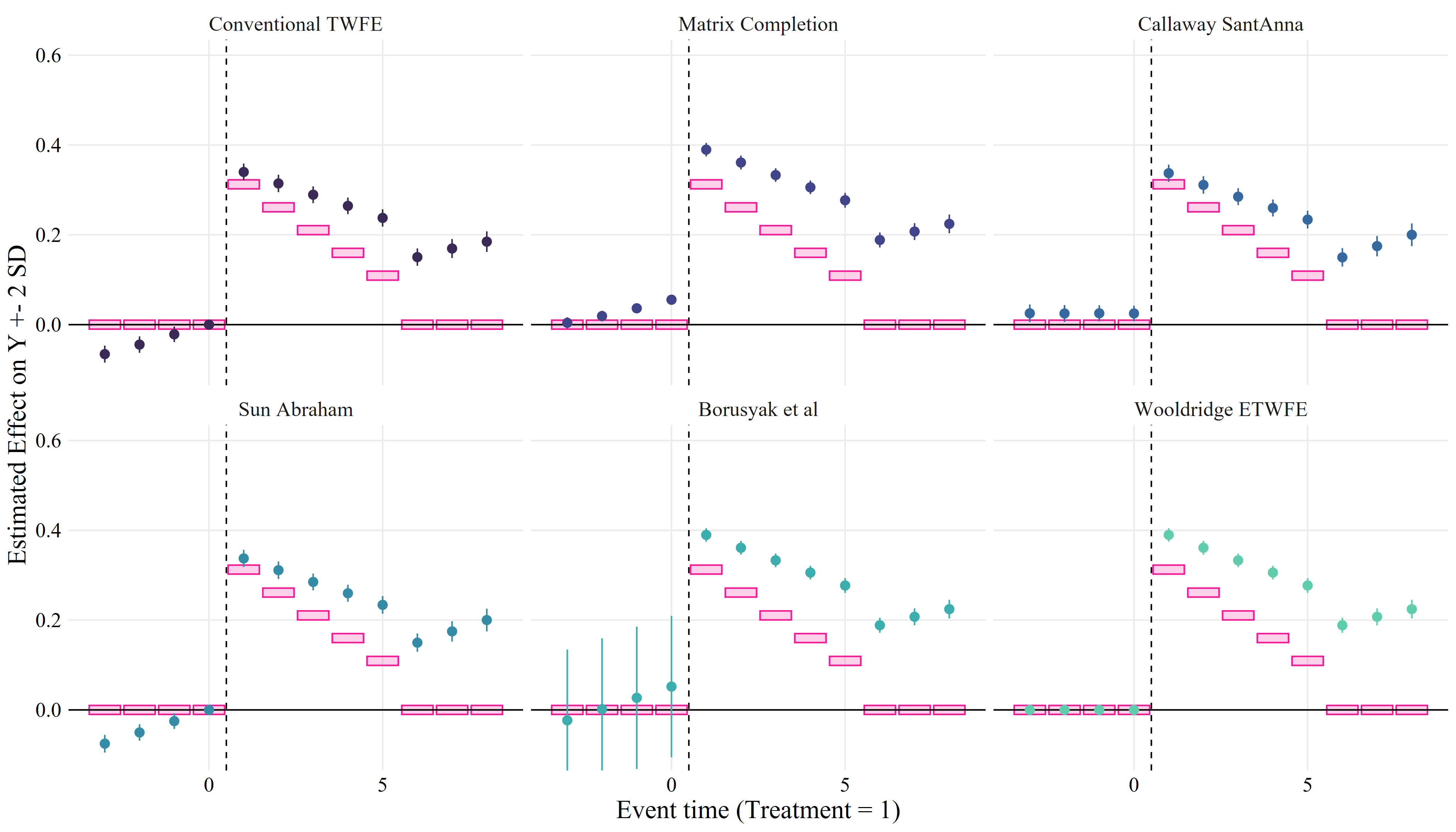}
    \caption{Alternative Set-up 6 Monte Carlo results: non-parallel trends, group-specific, time-heterogeneous (fading-out) treatment effects, no anticipation. Shown are the average coefficient estimates +/- two standard deviations. Pink rectangles mark the true effect according to the DGP.}
    \label{fig:fade-out6}
\end{figure}

\clearpage

\section{Two treatment timing groups} \label{suppl:twotime}

Below we changed the DGP in a way that treatment timing is not distributed across the 15 observation periods, but instead only happens at two distinct time periods. We thus have an early-treated group which is treated in period 4 and a late-treated group which is treated in period 12. As above, the early treated group experiences a larger treatment effect (twice as large as the late-treated).

\begin{figure}[h!]
    \includegraphics[width=\textwidth]{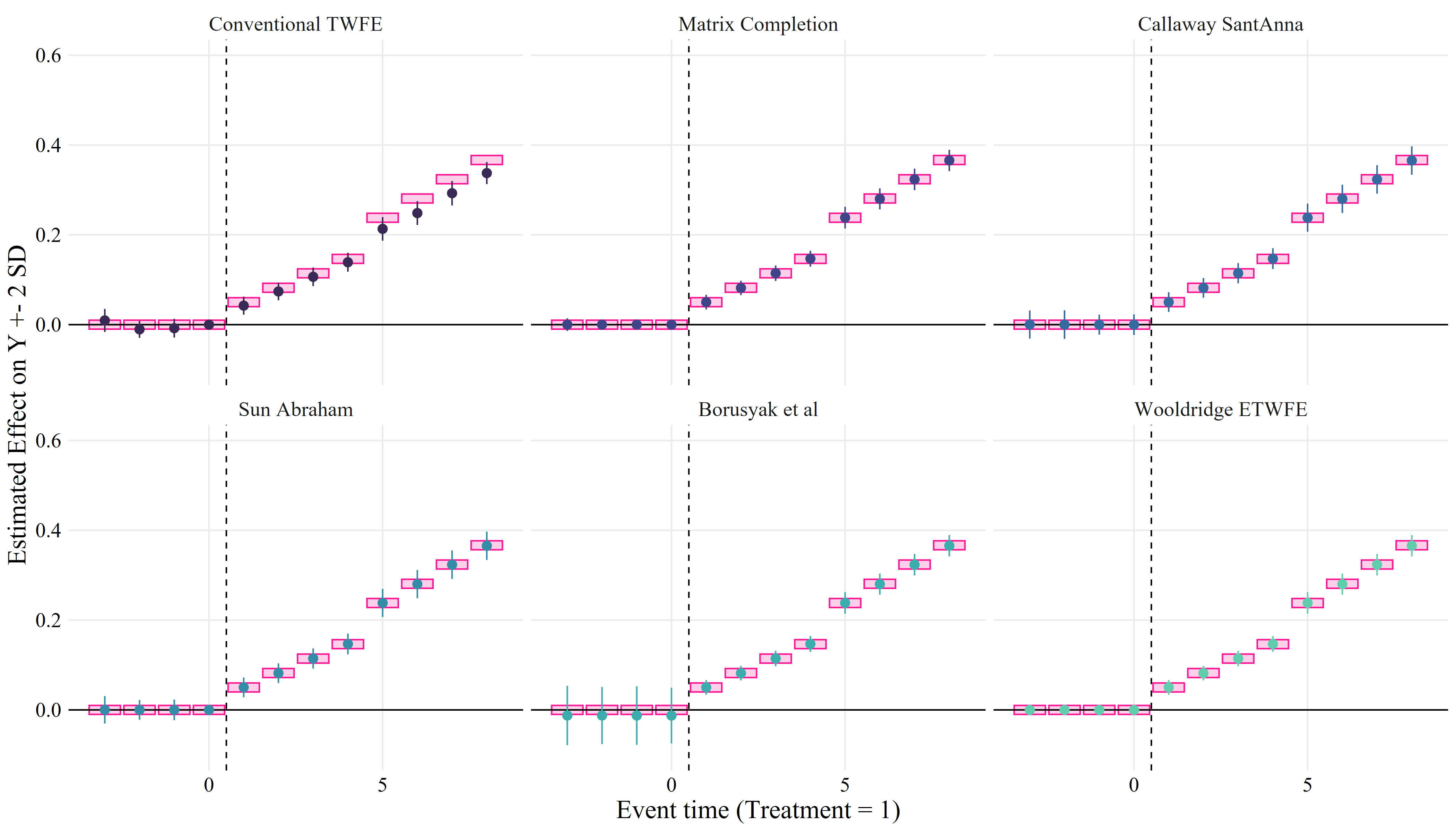} 
    \caption{Alternative Set-up 3 Monte Carlo results: group-specific and time-heterogeneous (trend-breaking) treatment effects with parallel trends and no anticipation. Instead of distributing the treatment timings across the observation window, we only define two treatment timings (early-treated in period 4, and late-treated in period 12). Shown are the average coefficient estimates +/- two standard deviations. Pink rectangles mark the true effect according to the DGP.}
    \label{fig:twotime1}
\end{figure}

\begin{figure}[h!]
    \includegraphics[width=\textwidth]{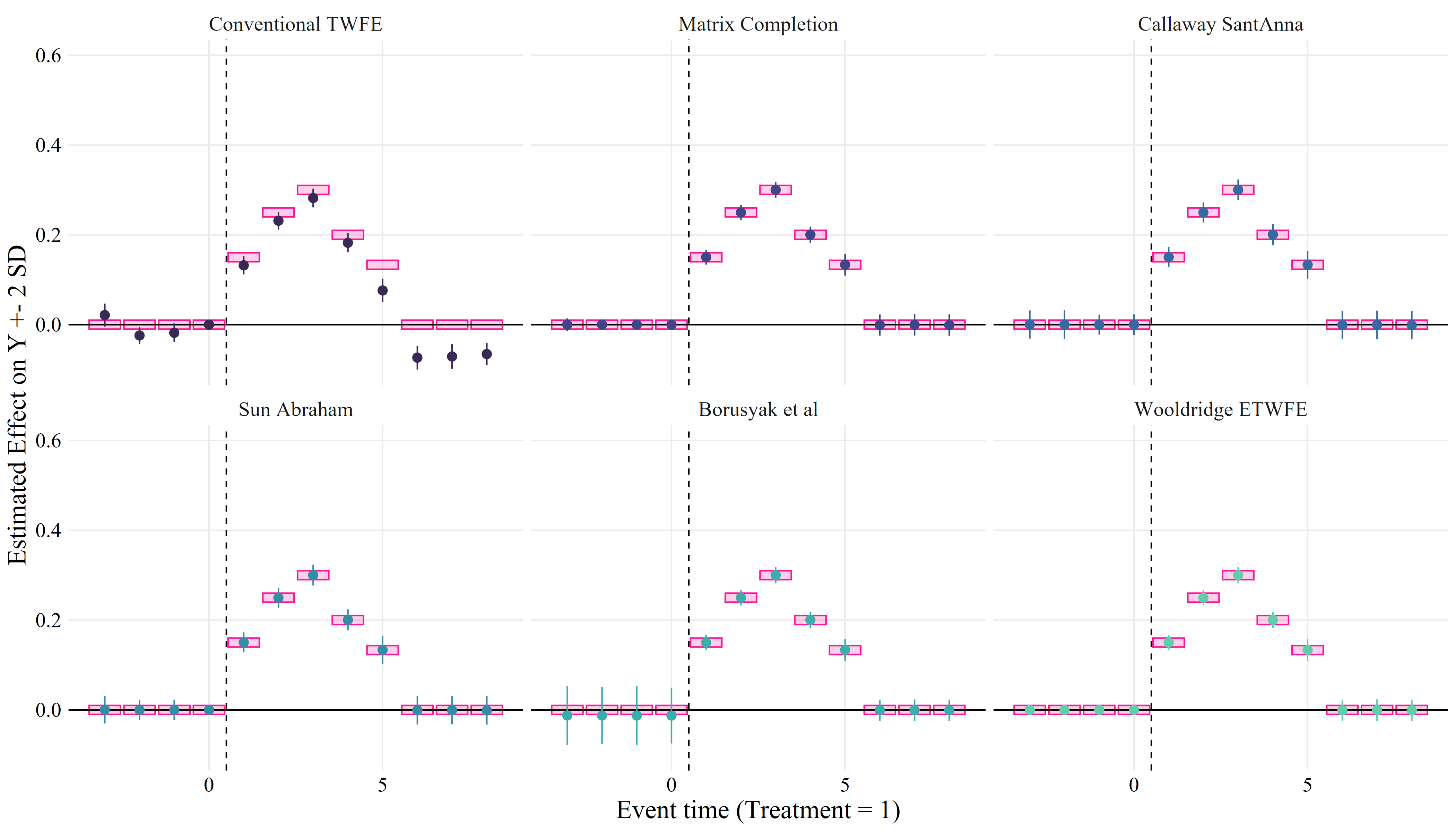} 
    \caption{Alternative Set-up 4 Monte Carlo results: parallel trends, group-specific, time-heterogeneous treatment effects (inverted-U). Instead of distributing the treatment timings across the observation window, we only define two treatment timings (early-treated in period 4, and late-treated in period 12). Shown are the average coefficient estimates +/- two standard deviations. Pink rectangles mark the true effect according to the DGP.}
    \label{fig:twotime1}
\end{figure}

\clearpage

\section{Small N, large T} \label{suppl:larget}

The DGP is identical to the main analysis, but we set $N = 300$, $T = 50$. Be aware of the extended x-axis due to large bias in panel 6.

\begin{figure}[h!]
    \centering
    \includegraphics[width=\textwidth]{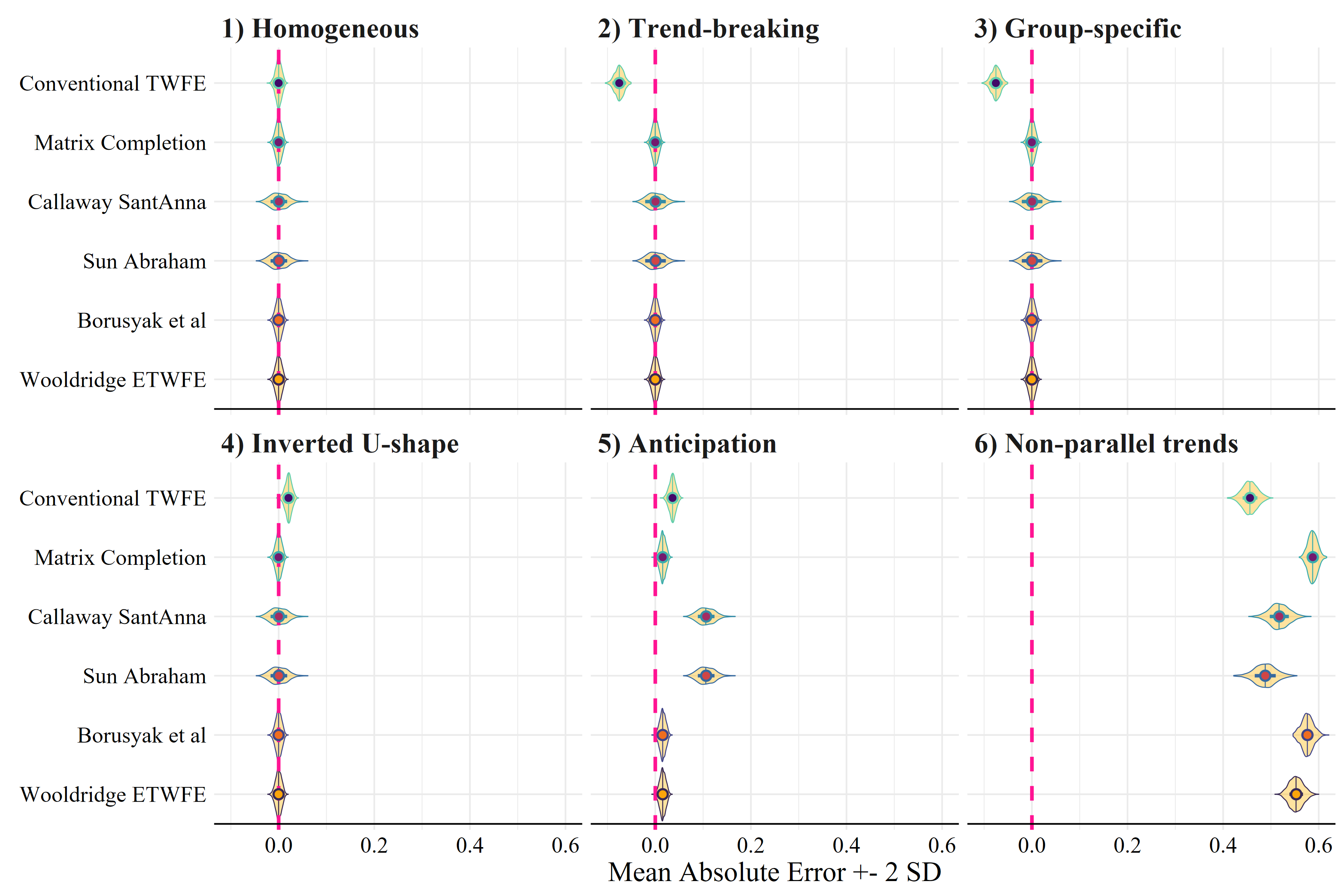}
    \caption{Monte Carlo results of single summary measure for the effect estimates. Set-ups 1 to 6 as described in Table \ref{tab:scenarios} but use a fading-out treatment form rather than an inverted u-shaped. Small N, large T setting with $N = 300$, $T = 50$. Shown are the average coefficient estimates +/- two standard deviations. The yellow violin graphs depict the distribution of the 1,000 individual estimates. The pink dashed line marks the true effect according to the DGP, centred around zero.}
    \label{fig:fig_att_larget}
\end{figure}

\end{document}